\pdfoutput=1
\documentclass[a4paper,12pt,margin=in]{article}
\usepackage{lineno}
\usepackage{natbib}
\usepackage[colorlinks,citecolor=blue,urlcolor=blue]{hyperref}
\usepackage{epsf}
\usepackage{epsfig}
\usepackage{multirow}
\usepackage[english]{babel}
\usepackage{graphicx}
\usepackage{geometry}
\usepackage{subcaption}
\usepackage{natbib}
\usepackage{amssymb}
\usepackage{amsmath}
\usepackage{color}
\usepackage{bm}
\usepackage{enumerate}
\usepackage{authblk}
\usepackage{float}
\usepackage{geometry}
 \usepackage{setspace}
\usepackage[font=footnotesize,labelfont=bf]{caption}

\newcommand{\specificthanks}[1]{\@fnsymbol{#1}}

\title{A novel family of beta mixture models for the differential analysis of DNA methylation data: an application to prostate cancer}
\author[1]{\normalsize{Koyel Majumdar}}
\author[2,3]{Romina Silva}
\author[3,4]{Antoinette Sabrina Perry}
\author[2,3]{Ronald William Watson}
\author[5]{Andrea Rau}
\author[5]{Florence Jaffrezic}
\author[1]{Thomas Brendan Murphy}
\author[1]{Isobel Claire Gormley\thanks{claire.gormley@ucd.ie}}
\affil[1]{\small{School of Mathematics and Statistics,University College Dublin, Ireland.}}
\affil[2]{School of Medicine, University College Dublin, Ireland.}
\affil[3]{Conway Institute of Biomedical and Biomolecular Research, University College Dublin, Ireland.}
\affil[4]{School of Biology and Environmental Science, University College Dublin, Ireland}
\affil[5]{INRAE, UMR1313 AgroParisTech, GABI, Université Paris-Saclay, France.}

\date{}

\begin{document}
\nolinenumbers
	\doublespacing
\maketitle

\begin{abstract}
Identifying differentially methylated cytosine-guanine dinucleotide (CpG) sites between benign and tumour samples can assist in understanding disease.  However, differential analysis of bounded DNA methylation data often requires data transformation, reducing biological interpretability. To address this, a family of beta mixture models (BMMs) is proposed that (i) objectively infers methylation state thresholds and (ii) identifies differentially methylated CpG sites (DMCs) given untransformed, beta-valued methylation data. The BMMs achieve this through model-based clustering of CpG sites and by employing parameter constraints, facilitating application to different study settings. Inference proceeds via an expectation-maximisation algorithm, with an approximate maximization step providing tractability and computational feasibility.

Performance of the BMMs is assessed through thorough simulation studies, and the BMMs are used for differential analyses of DNA methylation data from a prostate cancer study. Intuitive and biologically interpretable methylation state thresholds are inferred and DMCs are identified, including those related to genes such as GSTP1, RASSF1 and RARB, known for their role in prostate cancer development.  Gene ontology analysis of the DMCs revealed significant enrichment in cancer-related pathways, demonstrating the utility of BMMs to reveal biologically relevant insights. An R package \texttt{betaclust} facilitates widespread use of BMMs.

\end{abstract}

\section*{Introduction}
Epigenetics is the study of heritable changes in gene activity that do not involve explicit changes to the DNA sequence \citep{Berger}.
DNA methylation is an epigenetic process where a methyl group is added to or removed from the $5'$ carbon of the cytosine ring \citep{Moore}. This process assists in regulating gene expression and is essential for the development of an organism, but irregular changes in DNA methylation patterns can lead to damaging health effects \citep{Jin}. A cytosine-guanine dinucleotide (CpG) site is hypomethylated if neither of the DNA strands in a diploid individual 
are methylated, hemimethylated if either of the DNA strands are methylated or hypermethylated if both the strands are methylated. A differentially methylated CpG site (DMC) is a CpG site that has different methylation states between DNA samples collected from different biological conditions, which may have been taken from tissues of an individual over time, different tissues from the same individuals or distinct individuals.

The DNA methylation process has been extensively studied in the context of cancer, and its treatment \citep{Partha}.
CpG islands that remain unmethylated in normal cells can become methylated in abnormal cells such as cancer cells \citep{Bird}, and it has been shown that tumour suppressor genes are silenced by hypermethylation of their promoter regions \citep{Kim, Almeida}. For example, in prostate cancer, the fifth major cause of cancer-related mortality globally \citep{GCS2020}, hypermethylation of certain tumour suppressor genes, such as GSTP1, RARB, APC and RASSF1, has been observed during the early stages of the disease \citep{Longcheng, Moritz, Daniunaite}. A better understanding of disease can therefore be achieved by identifying regions that are differentially methylated between benign and tumour samples.

The Illumina MethylationEPIC BeadChip microarray \citep{Ruth} is used to interrogate over 850,000 CpG sites and retrieve methylation profiling of the CpG sites in the human genome. The Illumina microarray produces a sample of methylated ($Methylated$) and unmethylated ($Unmethylated$) light signal intensities, and the level of methylation, or the \textit{beta} value, is \textit{beta} = max(\textit{Methylated})/(max(\textit{Methylated}) + max(\textit{Unmethylated}) + $\chi$), where $\chi$ is a constant offset added for regularisation in case of very low \textit{Methylated} and \textit{Unmethylated} values \citep{Du}. The methylation level at a CpG site is quantified by this \textit{beta} value and is constrained to lie between $0$ and $1$. 
The \textit{beta} values are continuous with a value close to $1$ suggesting that a site is hypermethylated, while values close to $0$ represent hypomethylation. The two probe intensities are assumed to be gamma-distributed as they can take only positive values, and their ratio results in beta distributed variables. Thus, the \textit{beta} values can be modelled using a beta distribution.

The \textit{beta} values in general have higher variance in the center of the [0,1) interval than towards its endpoints. This leads to heteroscedasticity, which imposes challenges for analyses as assumptions for the ubiquitous Gaussian models are violated. Hence, \textit{beta} values are usually converted to \textit{M}-values using a logit transformation
as these values are statistically more convenient; 
 Gaussian models can be used as the transformed data lie within $(-\infty,\infty)$ \citep{Du}. However, such transformations make inference less biologically interpretable and hence there is a need to model the \textit{beta} values in their innate form. 

In many methylation array studies, thresholds of \textit{beta} values are subjectively selected to identify the three methylation states. For instance, \cite{ChenXin} deemed a CpG site to be hypomethylated if its \textit{beta} value was $<0.2$ and hypermethylated if its \textit{beta} value was $>0.8$, while \cite{men2017identification} employed $0.3$ and $0.7$ as thresholds. Such subjective selection of thresholds may increase the likelihood of false positives and negatives, leading to incorrect inference and necessitating an objective approach to determining  methylation state thresholds.

Mixture models for transformed \textit{beta} values have been proposed in several studies to find biologically meaningful clusters. For instance, \cite{Siegmund} and \cite{Koestler} use mixture models to model a subset of CpG sites and cluster samples into latent groups of biologically related samples. Additionally, methods such as the variational Bayes beta mixture model \citep{Zhanyu} and the Dirichlet process beta mixture model \citep{Lin} analyse untransformed \textit{beta} values; the former addresses the feature selection problem in the context of DNA methylation data, whereas the latter models the beta values to identify DNA methylation subgroups. The \texttt{Methylmix} \citep{Methylmix} R \citep{R2024} package uses a univariate beta mixture model to uncover patient subgroups with similar DNA methylation levels for a specific CpG site, with Wilcoxon rank sum tests used to establish hypermethylated and hypomethylated genes. 
Beta mixture models have also been proposed for intra-array quantile normalization \citep{teschendorff2013beta} and for clustering individual DNA samples into the three methylation states which can then be used to classify cancer tissue type \citep{laurila2011beta}. The use of a beta mixture model has been extended for classifying the methylation states of CpG sites; the approach accounts for boundary values and employs a method-of-moments approach to inference, but considers only small numbers of CpG sites \citep{schroder2017hybrid}. While mixture models for DNA methylation data have been used for a range of purposes, they have not been utilised to uncover differential methylation across the genome using untransformed \textit{beta} values.

Several methods have been developed for detecting DMCs in different DNA sample types. For instance, the \texttt{PanDM} method \citep{PanDM} leverages joint modeling to perform methylation site clustering, differential methylation detection, and pan-cancer pattern discovery by modelling the transformed p-values associated with each CpG site for a given cancer type. A principal component analysis and tensor decomposition approach involving unsupervised feature selection \citep{PCATD} was proposed where principal component scores were associated with each CpG site and used to identify DMCs. Another approach, termed \texttt{FastDMA} \citep{FastDMA}, employs an analysis of covariance to perform both single probe analysis and differentially methylated region scanning while modelling the $M$-values. The popular \texttt{limma} method \citep{limma} identifies DMCs by modelling the \textit{M}-values using an empirical Bayesian approach. Other studies identify the DMCs by modelling the \textit{beta} values via multiple moderated t-tests or Wilcoxon rank-sum tests  \citep{IMA,COHCAP}. Additionally, a multiple hypothesis testing approach, combined with multivariate permutation tests, has been proposed to detect group differences in epigenetic data \citep{Olek}, as has a nonparametric test to identify DMCs between multiple treatments \citep{Zhongxue}; while this approach can analyze smaller arrays with e.g., 28,000 CpG sites, it is computationally intensive for modern, larger-scale arrays.
These approaches to DMC identification use subjective thresholds, transformed values, moderated $t$-tests and/or nonparameteric methods \citep{Wang}. Crucially, such approaches lack biological interpretability and often face reproducibility and computational scalability challenges when considering data from different studies, of the scale resulting from current microarray technologies.

Several mixture models for bounded data are available. For instance, in the context of semiparametric density estimation, \cite{GMDEB} fit a Gaussian mixture model to range-power transformed bounded data, from which the density for the original data is obtained. Mixture models of bounded Laplace distributions also allow for modelling bounded data by truncation of the Laplace distribution, but are computationally expensive for large datasets \citep{BLMM}. Similarly, bounded support asymmetric generalized Gaussian mixture models are adaptable to different distributional shapes but can be computationally expensive as inference requires numerical optimisation \citep{BAGGMM}. A beta mixture model is an appropriate choice for bounded DNA methylation data: the support of the beta distribution is congruent with the \textit{beta} values, its flexibility allows for skew and symmetric distributional shapes, and it is computationally feasible to work with given its parsimony. Importantly, the beta distribution parameters provide relevant biological interpretations enabling biologically intuitive and meaningful inference.

Here we propose a family of beta mixture models (BMMs), which address specific research questions arising in the context of differential analysis of DNA methylation data, by introducing a range of constraints on the parameters of a BMM. The resulting novel family of BMMs facilitates a model-based approach to clustering CpG sites given their innate beta-valued methylation data to (i) objectively identify methylation state thresholds and (ii) identify DMCs between different sample types.
The BMMs are capable of clustering the entire microarray of CpG sites, from DNA samples collected from multiple tissues from each of several patients, in a computationally efficient manner. Performance is assessed through simulation studies, and the BMMs are used to analyse a motivating prostate cancer (PCa) dataset. The capability of the BMMs is demonstrated to appositely model the \textit{beta} values, 
to objectively identify thresholds and to identify existing and novel DMCs, including those related to genes implicated in prostate cancer, such as GSTP1, RARB and RASSF1. An R package, \texttt{betaclust}, freely available on  \href{https://github.com/koyelucd/betaclust}{github} and \href{https://cran.r-project.org/web/packages/betaclust/index.html}{CRAN}, facilitates widespread use of the BMMs.

\section*{Methods}
\subsection*{Prostate cancer data}
A prostate cancer study \citep{Silva}, which involved collection of DNA methylation samples from four patients with metastatic prostate cancer disease, motivated the development of the BMMs.
Tissue samples from matched biopsy cores (tumour and histologically matched normal – herein benign) were collected from each patient, and DNA was extracted from the samples. Methylation profiling of the DNA samples was conducted using the Infinium MethylationEPIC Beadchip \citep{EPIC}. The raw DNA methylation data are freely available for download on the Gene Expression
Omnibus (GEO) repository (GSE119260); datasets GSM3362390-GSM3362397 were analysed here and were accessed
on 26\textsuperscript{th} of January, 2021 for research purposes. The authors had no access to information that could identify individual participants.

Observed \textit{beta} values for each of 694,923 CpG sites for the two DNA sample types were collected from each of the four patients. Raw methylation array data was quality controlled and pre-processed as in
\cite{Silva}, where the data were normalized, and probes overlapping with SNPs, probes with the highest fraction of unreliable measurements, probes lying outside of CpG sites and those on the sex chromosome were removed. The resulting dataset had 103 CpG sites ($<$ 0.014\% of the total number of CpG sites) with missing \textit{beta} values. While imputation techniques exist for DNA methylation data, missing values were not imputed here due to their very low percentage, and the high uncertainty associated with imputed values in diseased samples due to their heterogeneity \citep{Lena}; here the 103 CpG sites with missing data
were therefore removed. No observed \textit{beta} values were equal to $0$.
The resulting dataset contained \textit{beta} values for $C = 694,820$ CpG sites from each of $R = 2$ DNA sample types collected from each of $N = 4$ patients.  Here, these data are appositely modeled in their innate \textit{beta} form to  (i) objectively identify methylation state thresholds and (ii) uncover DMCs between two sample types using a model-based clustering approach.

\subsection*{\textbf{A beta distribution}}
The beta distribution has support on $[0,1]$ and is parameterized by two positive shape parameters, $\alpha$ and $\delta$. 
Given the properties of the \emph{beta} values, the beta distribution is used here to appositely model the methylation level $x_{cnr}$ of the $c^{th}$ CpG site ($c = 1,\ldots,C)$, from the $n^{th}$ patient ($n = 1,\ldots,N)$, from their $r^{th}$ DNA sample type ($r = 1,\ldots,R$) i.e., 
\[
    f(x_{cnr}|\alpha,\delta)  \sim \mbox{Beta}({x_{cnr}}|\alpha, \delta) = \frac{{x_{cnr}}^{\alpha-1}(1-{x_{cnr}})^{\delta-1}}{\mbox{B}(\alpha,\delta)},  
\] 
for $0 \leq  x_{cnr}\leq 1$, where $\mathrm{B}(\alpha,\delta) = (\Gamma(\alpha)\Gamma(\delta))/\Gamma(\alpha+\delta)$, where $\Gamma(\cdot)$ is the gamma function. The DNA methylation data are collected in the  $C\times NR$ dimensional dataset $\mathbf{X}$ where each of the $NR$ columns contains the methylation levels of the $C$ CpG sites in one of the $R$ sample types from each of the $N$ patients.
\newline

\subsection*{\textbf{A beta mixture model}}
A mixture model assumes the observed data have been generated from a heterogeneous population composed of $K$ groups or clusters. In the context of DNA methylation data, there are $G = 3$ possible methylation states: hypomethylation, hemimethylation or hypermethylation. Hence, when analysing methylation data from a single DNA sample type (i.e., where $R = 1$) each CpG site exhibits one of $K = G^R = 3$ methylation states characterised by each of the $K$ clusters. Here, interest lies in objectively inferring thresholds between these $K = 3$ methylation states.

When analysing methylation data across multiple (i.e., $R > 1$) DNA sample types to identify DMCs, each CpG site will exhibit one of a possible $K = G^R$  combinations of methylation states, here characterised by each of $K$ clusters in a mixture model. For example, given the three methylation states and considering a CpG site across $R = 2$ sample types (e.g., across benign and tumour samples), the CpG site can potentially exhibit any of the $K = 3^2 = 9$ combinations of these three states (for example hypermethylated in both samples, hypermethylated in one sample and hypomethylated in the other, etc). Therefore, in this scenario, $K = 3^2 = 9$ with each cluster characterising one of the possible methylation state combinations.

We propose a beta mixture model for the methylation data for all cases $R \ge 1$.
Here $\bm{\theta}$ is used to denote the shape parameters in a beta mixture model, i.e., $\bm{\theta} = (\bm{\alpha}_1,\bm{\delta}_1,...,\bm{\alpha}_K,\bm{\delta}_K)$, where $\bm{\alpha}_k$ and $\bm{\delta}_k$ are the shape parameters of cluster $k$. The shape parameters are allowed to vary among the clusters, patients and sample types. The mixing proportions $\bm{\tau} = $($\tau_1$,...,$\tau_K$) lie between $0$ and $1$, $\sum_{k=1}^{K} \tau_{k} = 1$, and denote the probability of belonging to cluster $k$ $\forall   k = 1,\ldots,K$. Independence is assumed across patients and samples, given a CpG site's cluster membership, leading to the probability density function for such a beta mixture model (BMM):

\begin{equation}
\label{equation 6a}
    f(\mathbf{X}|\bm{\tau},\bm{\theta}) =  \prod_{c=1}^{C} \sum\limits_{k=1}^{K} \tau_k \textit{f} (\mathbf{X}|\bm{\alpha}_k,\bm{\delta}_k) 
     = \prod_{c=1}^{C} \sum\limits_{k=1}^{K} \tau_k \prod_{n=1}^{N} \prod_{r=1}^{R} \mbox{Beta}(x_{cnr}|\alpha_{knr},\delta_{knr})
\end{equation}

Computation of maximum likelihood estimates (MLEs) of $\bm{\tau}$  and $\bm{\theta}$ from 
the associated log likelihood function is complex, and an incomplete data approach is therefore used here. The latent vector $\mathbf{z}_c = (z_{c1}, \ldots,z_{cK})$ is introduced for each CpG site $c$, where $z_{ck}$ is $1$ if CpG site $c$ belongs to the $k^{th}$ group and $0$ otherwise. The $C \times K$ matrix $\mathbf{Z}$ is combined with the \textit{beta} values to form the complete data $(\mathbf{X},\mathbf{Z})$. The complete data log-likelihood function is
\begin{equation}
\label{equation 10a}
  \ell_{C}(\bm{\tau}, \bm{\theta},\mathbf{Z}|\mathbf{X}) = \sum\limits_{c=1}^{C}\sum\limits_{k=1}^{K} z_{ck} \{\log  \tau_k + 
  \sum\limits_{n=1}^{N} \sum\limits_{r=1}^{R}
  \log [\mbox{Beta}(x_{cnr}|\alpha_{knr},\delta_{knr})]\}.
\end{equation}
The complete data log-likelihood function (\ref{equation 10a}) can be used to find the MLEs $\hat{\bm{\tau}}$ and $\hat{\bm{\theta}}$ using the expectation-maximisation (EM) algorithm \citep{EM};  on convergence  a probabilistic clustering solution is also available from the expected value of $z_{ck}$, the posterior probability of CpG site $c$ belonging to cluster $k$.  

\subsubsection*{\textbf{A family of BMMs}}
The most generalised BMM is defined in (\ref{equation 6a}) which models the CpG sites as belonging to $K$ latent groups. By introducing a variety of constraints on the parameters of this generalised BMM, a family of three beta mixture models is proposed. Each model serves a specific purpose e.g.,  to cluster the CpG sites into the 3 methylation states allowing objective inference of methylation state thresholds, or to facilitate the identification of DMCs between different sample types.

\paragraph{\textbf{The K$\cdot\cdot$ model}}
The K$\cdot\cdot$ model facilitates objective inference of thresholds between methylation states by clustering $C$ CpG sites into one of $K = G = 3$ methylation states, based on a single sample type ($R = 1$) from each of $N$ patients. Under the K$\cdot\cdot$ model the shape parameters of each cluster are constrained to be equal for each patient, but allowed to vary across clusters. The complete data log-likelihood function is therefore
\[
  \ell_{C}(\bm{\tau}, \bm{\theta},\mathbf{Z}|\mathbf{X}) = \sum\limits_{c=1}^{C}\sum\limits_{k=1}^{K} z_{ck} \{\log  \tau_k + 
  \sum\limits_{n=1}^{N} \sum\limits_{r=1}^{1}
   \log [\mbox{Beta}(x_{cnr}|\alpha_{k\cdot\cdot},\delta_{k\cdot\cdot})]\}.
\]

\paragraph{\textbf{The KN$\cdot$ model}}
The  KN$\cdot$ model facilitates objective inference of methylation state thresholds by clustering each of the $C$ CpG sites into one of $K = G = 3$ methylation states, based on data from a single sample type ($R = 1$) from each of $N$ patients. While the KN$\cdot$ model has a similar purpose to the K$\cdot\cdot$ model, it  differs in that it is less parsimonious as it allows cluster and patient-specific shape parameters. The complete data log-likelihood function is therefore
\[
  \ell_{C}(\bm{\tau},\bm{\theta},\mathbf{Z}|\mathbf{X}) = \sum\limits_{c=1}^{C}\sum\limits_{k=1}^{K} z_{ck} \{\log  \tau_k +  
  \sum\limits_{n=1}^{N} \sum\limits_{r=1}^{1}
   \log [\mbox{Beta}(x_{cnr}|\alpha_{kn\cdot},\delta_{kn\cdot})]\}.
\]

\paragraph{\textbf{The K$\cdot$R model}}
The K$\cdot$R model facilitates identification of differentially methylated CpG sites between $R > 1$ DNA sample types collected from each of $N$ patients. The K$\cdot$R model assumes conditional independence between CpG sites from paired samples from the same patient, given the CpG sites' cluster membership. The K$\cdot$R model also assumes each of the $K$ clusters characterises a different combination of the $G$ methylation states across the $R$ biological conditions where $K = G^R = 9$ here. 
Under the K$\cdot$R model the shape parameters are allowed to vary for each sample type and for different clusters but are constrained to be equal for each patient. The complete data log-likelihood function for the K$\cdot$R model is therefore
 \[
  \ell_{C}(\bm{\tau}, \bm{\theta},\mathbf{Z}|\mathbf{X}) = \sum\limits_{c=1}^{C}\sum\limits_{k=1}^{K} z_{ck} \{\log  \tau_k + \sum\limits_{n=1}^{N} \sum\limits_{r=1}^{R}
   \log [\mbox{Beta}(x_{cnr}|\alpha_{k \cdot r},\delta_{k \cdot r})]\}.
 \]

This family of beta mixture models enables the objective inference of methylation state thresholds (via the K$\cdot\cdot$ and/or KN$\cdot$ models) and the identification of DMCs between $R$ DNA sample types (via the K$\cdot$R model), as illustrated in the simulation studies and applications that follow.

\subsection*{\textbf{Parameter estimation}}
The parameters of the BMMs are estimated and the cluster membership for each CpG site inferred using the EM algorithm. Here, we delineate this for the generalised BMM (\ref{equation 6a}). Derivations for the K$\cdot\cdot$, KN$\cdot$ and K$\cdot$R models are detailed in Appendices S1--S3.

The EM algorithm consists of two steps: in the expectation step the expected value of the complete data log-likelihood function is obtained, conditional on the observed data and current parameter estimates. The maximisation step maximises the expected complete data log-likelihood with respect to the parameters. To obtain $\bm{\hat{\tau}}$ and $\bm{\hat{\theta}}$, the expectation and maximisation steps are iterated  until convergence to at least a local optimum of the log-likelihood function.

An initial clustering of CpG sites is obtained using k-means clustering and the method of moments is used to calculate initial values of $\bm{\tau}$ and $\bm{\theta}$. The two steps proceed as follows:

\begin{itemize}
    \item Expectation-step: the expected value of $z_{ck}$ is calculated, i.e., the posterior probability of CpG site $c$ belonging to cluster $k$, conditional on current parameter estimates. At iteration $t+1$  
\[
        \hat{z}_{ck} =   \textbf{E}[z_{ck}|\mathbf{X}, \bm{\tau}^{(t)}, \bm{\theta}^{(t)}] 
        =  \frac{\tau_k^{(t)} \prod_{n=1}^{N} \prod_{r=1}^{R} \mbox{Beta}(x_{cnr}|\alpha_{knr}^{(t)},\delta_{knr}^{(t)})}{\sum\limits_{k'=1}^K \left[\tau_{k'}^{(t)} \prod_{n=1}^{N}\prod_{r=1}^{R} \mbox{Beta}(x_{cnr}|\alpha_{k'nr}^{(t)},\delta_{k'nr}^{(t)})\right] }.
\]
    \item Maximisation-step: estimates of the parameters $\bm{\tau}$ and $\bm{\theta}$ are calculated by maximising the expected complete data log-likelihood function, given the $\mathbf{\hat{Z}}$ values from the E-step. 
\end{itemize}
For the maximisation-step, the expected complete data log-likelihood function is maximised by differentiating it w.r.t the parameters. Closed form solutions for the mixing proportions are available as
$    \hat{\tau}_k = \sum\limits_{c=1}^C \hat{z}_{ck}/C,\;   \forall \; k = 1,\ldots, .K.
$
For the shape parameters, the expected complete data log-likelihood function to be maximized is
\begin{equation}
\label{equation 14a}
\begin{aligned}
    \ell_{C}(\bm{\tau}, \bm{\theta}|\mathbf{X},\mathbf{\hat{Z}}) = &\sum\limits_{c=1}^C \sum\limits_{k=1}^K \hat{z}_{ck} \{ \log \tau_k + \sum\limits_{n=1}^N\sum\limits_{r=1}^{R}[(\alpha_{knr}-1)\log x_{cnr} + \\
    &(\delta_{knr}-1)\log (1-x_{cnr})- 
    \log \mbox{B}(\alpha_{knr},\delta_{knr})]\}.
    \end{aligned}
\end{equation}
Differentiating (\ref{equation 14a}) w.r.t $\alpha_{knr}$ yields
\begin{equation}
\label{equation 15a}
    \frac{\partial{\ell_{C}}}{\partial{\alpha_{knr}}} =  \sum\limits_{c=1}^C \hat{z}_{ck} \{\log x_{cnr} - 
    [\psi(\alpha_{knr})-
    \psi(\alpha_{knr}+\delta_{knr})]\}
\end{equation}
where $\psi$ is the logarithmic derivative of the gamma function known as the digamma function,    $
    \psi(\alpha_{knr}) = {\partial{\log \Gamma(\alpha_{knr})}}/{\partial \alpha_{knr}}$. Similarly, the derivative of $\ell_{C}(\bm{\tau}, \bm{\theta}|\mathbf{X},\mathbf{\hat{Z}})$ \newline w.r.t $\delta_{knr}$ is
\begin{equation}
\label{equation 17}
    \frac{\partial{\ell_{C}}}{\partial{\delta_{knr}}} = \sum\limits_{c=1}^C \hat{z}_{ck} \{\log (1-x_{cnr}) - [\psi(\delta_{knr})-\psi(\alpha_{knr}+\delta_{knr})]\}.
\end{equation}
Closed form solutions for $\hat{\alpha}_{knr}$ and $\hat{\delta}_{knr}$ are not available due to the presence of the digamma function. To obtain the MLEs, numerical optimisation algorithms such as BFGS \citep{BFGS} and BHHH \citep{BHHH} could be used. However, for the large datasets considered here, use of these algorithms proved to be computationally infeasible.

\subsubsection*{\textbf{A digamma approximation}}
Here an approximation to the digamma function is used to allow for closed form solutions for the shape parameters. 
The lower bound for the digamma function for all  $y > 1/2$ is $\psi(y) > \log(y-1/2)$ \citep{Straub}. Given the context, we assume that the beta distributions in the family of BMMs are unimodal and bounded, meaning the shape parameters are $> 1$. Thus, the lower bound approximation holds and was empirically observed to be a very close approximation of the digamma function. The lower bound is used in (\ref{equation 15a}) and (\ref{equation 17}) to give
\begin{equation}
\label{equation 21}
    \frac{\partial{\ell_{C}}}{\partial{\alpha_{knr}}} \approx  \sum\limits_{c=1}^C \hat{z}_{ck} \sum\limits_{n=1}^N \sum\limits_{r=1}^{R} \biggl[ \log x_{cnr} -    \log \frac{\alpha_{knr}-1/2}{\alpha_{knr}+\delta_{knr}-1/2} \biggr]
\end{equation}
and
\begin{equation}
\label{equation 22}
    \frac{\partial{\ell_{C}}}{\partial{\delta_{knr}}} \approx \sum\limits_{c=1}^C \hat{z}_{ck} \sum\limits_{n=1}^N \sum\limits_{r=1}^{R}\biggl[\log (1-x_{cnr}) - \log \frac{\delta_{knr}-1/2}{\alpha_{knr}+\delta_{knr}-1/2}\biggr].
\end{equation}
Equating equations (\ref{equation 21}) and (\ref{equation 22}) to zero, we get closed-form, approximate estimates as
\[        \hat{\alpha}_{knr} = 0.5+ \frac{0.5\exp (-y_2)}{\{[\exp(-y_2)-1][\exp(-y_1)-1]\}-1}\]
and
\[        \hat{\delta}_{knr} =  \frac{0.5\exp (-y_2)[\exp(-y_1)-1]}{\{[\exp(-y_2)-1][\exp(-y_1)-1]\}-1},\]
where
        $y_1 = (\sum\limits_{c=1}^C \hat{z}_{ck} \log x_{cnr})/(\sum\limits_{c=1}^C \hat{z}_{ck})$ and $y_2 = (\sum\limits_{c=1}^C \hat{z}_{ck} \log (1-x_{cnr}))/(\sum\limits_{c=1}^C \hat{z}_{ck})$.

Utilising the digamma function approximation brings notable computational gains with run times, for example, reducing from 65 hours (when using numerical optimisation at the maximisation step) to 15 minutes (when using the digamma approximation) when analysing the PCa data on a computer equipped with an Intel Core i7 CPU with 2.70GHz speed, 6 physical cores and 16 GB of RAM.

\subsection*{\textbf{Inferring methylation state thresholds}}
To objectively infer thresholds between methylation states, without loss of generality, we denote by clusters 1 and 2 the clusters representing hypomethylated and hypermethylated CpG sites respectively. The ratio of fitted density estimates $\omega_j$ for cluster $j = 1, 2$ is
\[
 \omega_j = \frac{\tau_{j} \textit{f} (\mathbf{X}|\bm{\alpha}_{j} ,\bm{\delta}_{j} ) }{\sum\limits_{k\neq {j}} \tau_k \textit{f} (\mathbf{X}| \bm{\alpha}_{k},\bm{\delta}_k)} .
\]
The threshold separating e.g., the hypomethylated and hemimethylated clusters is calculated as the minimum \textit{beta} value at which $\omega_1 \ge 1$. Similarly, the threshold dividing the hemimethylated and hypermethylated clusters is the maximum \textit{beta} value at which $\omega_2 \ge 1$. 

In the K$\cdot\cdot$ model, as the shape parameters are constrained to be equal for each patient, a single set of thresholds is calculated for all patients. In the KN$\cdot$ model, the shape parameters vary for each patient, so a set of thresholds is calculated for each patient. Unless one model is appropriate given the question of interest, to choose the optimal model between the K$\cdot\cdot$ and KN$\cdot$ models, here the well utilised model selection tools of the Akaike information criterion (AIC) \citep{AIC}, Bayesian information criterion (BIC) \citep{BIC} and the integrated complete log-likelihood criterion (ICL) \citep{Biernacki} are examined.

\subsection*{\textbf{Identifying the most differentially methylated clusters}}
 While the K$\cdot$R model clusters the CpG sites into $K$ clusters, subsequent quantification of the degree of differential methylation of CpG sites in each cluster is required. Here, this is quantified by comparing the $R = 2$ beta distributions associated with the two sample types in each cluster (e.g., benign and tumour) using the area under the curve (AUC) of the receiver operating characteristic (ROC) curve as a measure of separability, which has an advantage of being widely used in biology. 
 
The ROC curve for a cluster is generated by drawing 1000 samples from each of the two fitted beta distributions, corresponding to two sample types, after convergence of the EM algorithm. By varying threshold values from 0 to 1, the sampled \textit{beta} values are used to compute sensitivity and specificity for discriminating between sample types, which are in turn used to construct the ROC curve and associated AUC. Higher AUC values indicate greater separation between the fitted beta distributions for two sample types within a cluster, and therefore that the CpG sites within that cluster are more differentially expressed between the two. In the case where $R > 2$, the beta distributions linked to each sample type within a cluster are compared to one another, and the maximum AUC across pairs of sample types is selected as the cluster's AUC. A large AUC would then indicate distinct methylation patterns between at least two sample types.
We also consider the Wasserstein distance (WD) \citep{WD}, which computes the disparity between cumulative distributions, as an additional approach to quantifying the degree of differential methylation of CpG sites between sample types in each cluster. 

\begin{figure}[h!]
\begin{center}
\includegraphics[width=\textwidth,height=5cm]{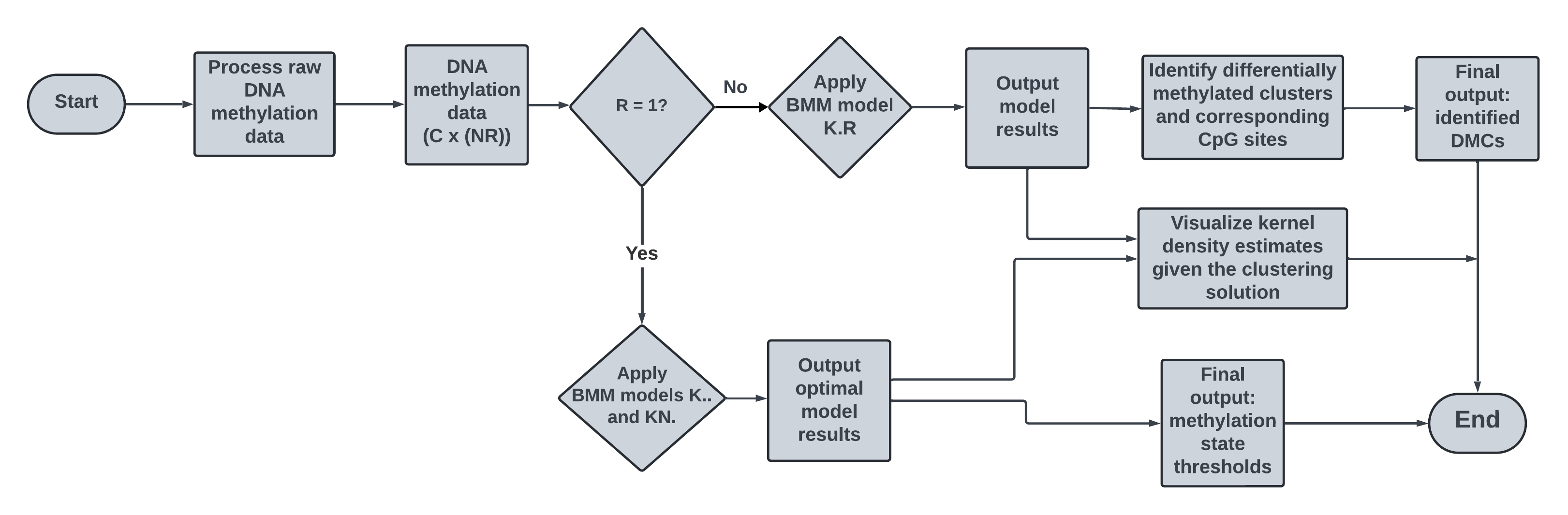}
\caption{
\textbf{Process flow diagram for the differential analysis of beta-valued DNA methylation data using a novel family of beta mixture models.}}

\label{Fig1}
\end{center}
\end{figure}

A process flow diagram that summarises the overall approach to inferring subjective thresholds and identifying DMCs using the proposed BMMs is given in Fig \ref{Fig1}.

\section*{Results}

\subsection*{\textbf{Simulated data results}}
\subsubsection*{\textbf{Simulated data}}
\begin{figure}[h!]
\begin{center}
\includegraphics[width=\textwidth,height=0.7\textheight]{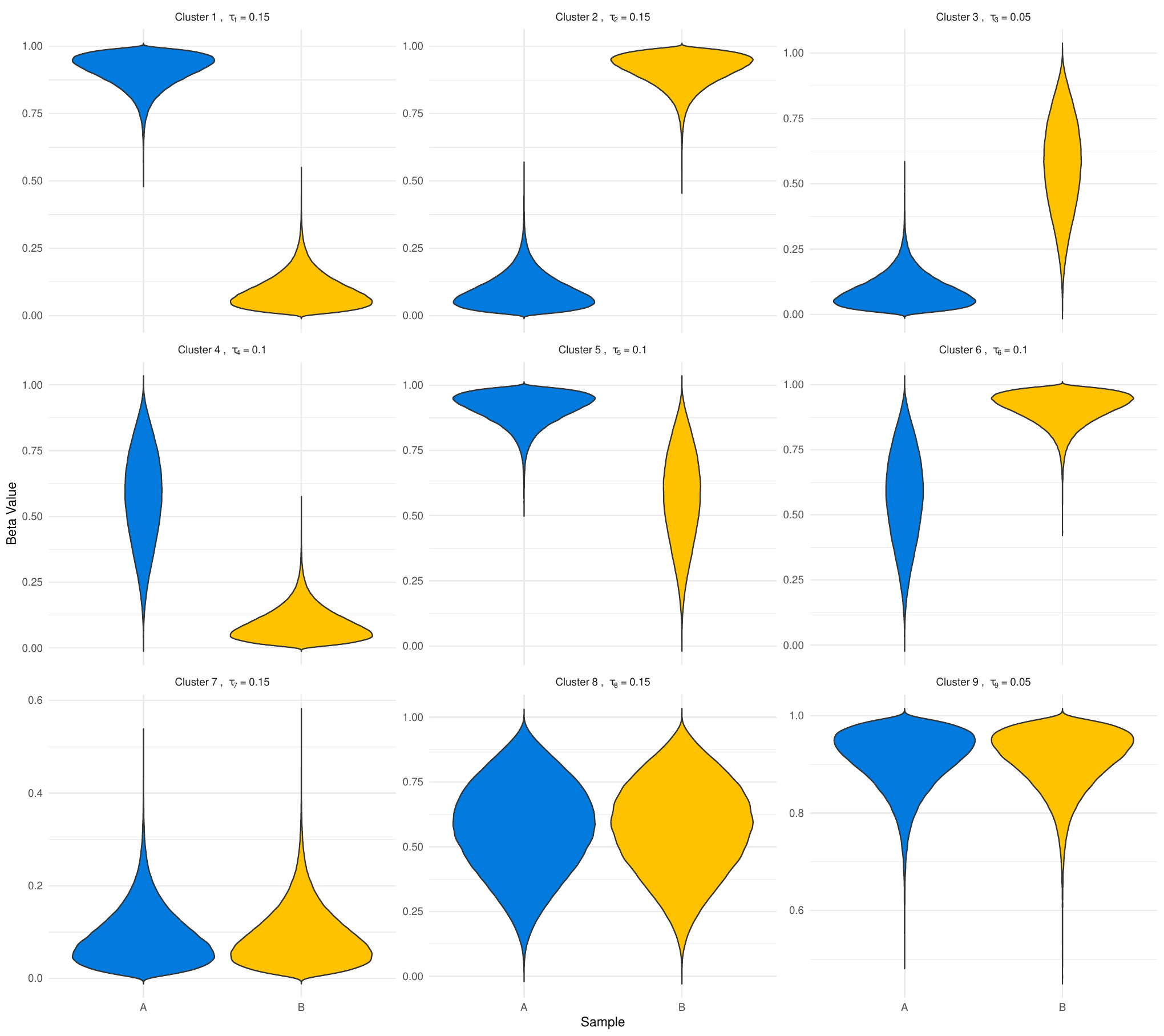}
\caption{\textbf{Violin plots of a simulated dataset.}}
\caption*{\small{Each panel illustrates the simulated beta-distributed values in samples A and B. The proportion of CpG sites in each cluster is detailed in the panel title. Clusters are ordered numerically from most to least differentially methylated, according to AUC.}}
\label{Fig2}
\end{center}
\end{figure}

One hundred simulated datasets consisting of methylation values for $C = 600,000$ CpG sites are generated using R \citep{R2024}.
Each simulated dataset consists of \emph{beta} values from two biological sample types (sample A and sample B) from each of $N = 4$ patients. Hypomethylated CpG sites were generated from a \textit{Beta}(2,20) distribution, with hemi- and hyper- values generated from \textit{Beta}(4,3) and \textit{Beta}(20,2) distributions respectively. To emulate the noisy data observed in real settings, zero-centred Gaussian noise with standard deviation 0.01 was added to the beta-generated data; resulting values outside $[0, 1]$ were replaced by the closest minimum or maximum value from the beta-generated data. Reflecting typical behaviour in DNA methylation data, 35\%, 35\% and 30\% of CpG sites in a single sample were simulated as hypomethylated, hemimethylated and hypermethylated, respectively. This resulted in, on average, 64\% of the CpG sites being differentially methylated between the two sample types.
Of these DMCs, on average, 30\% were hypomethylated in one sample and hypermethylated in the other, or vice versa; such highly differentially methylated CpG sites are of prime interest. Fig \ref{Fig2} illustrates a single simulated data set, with clusters of CpG sites ordered from most to least differentially methylated between samples, according to AUC.

\subsubsection*{\textbf{Estimating methylation state thresholds}}
\begin{figure}
\begin{center}
\includegraphics[width=0.8\textwidth,height=9cm]{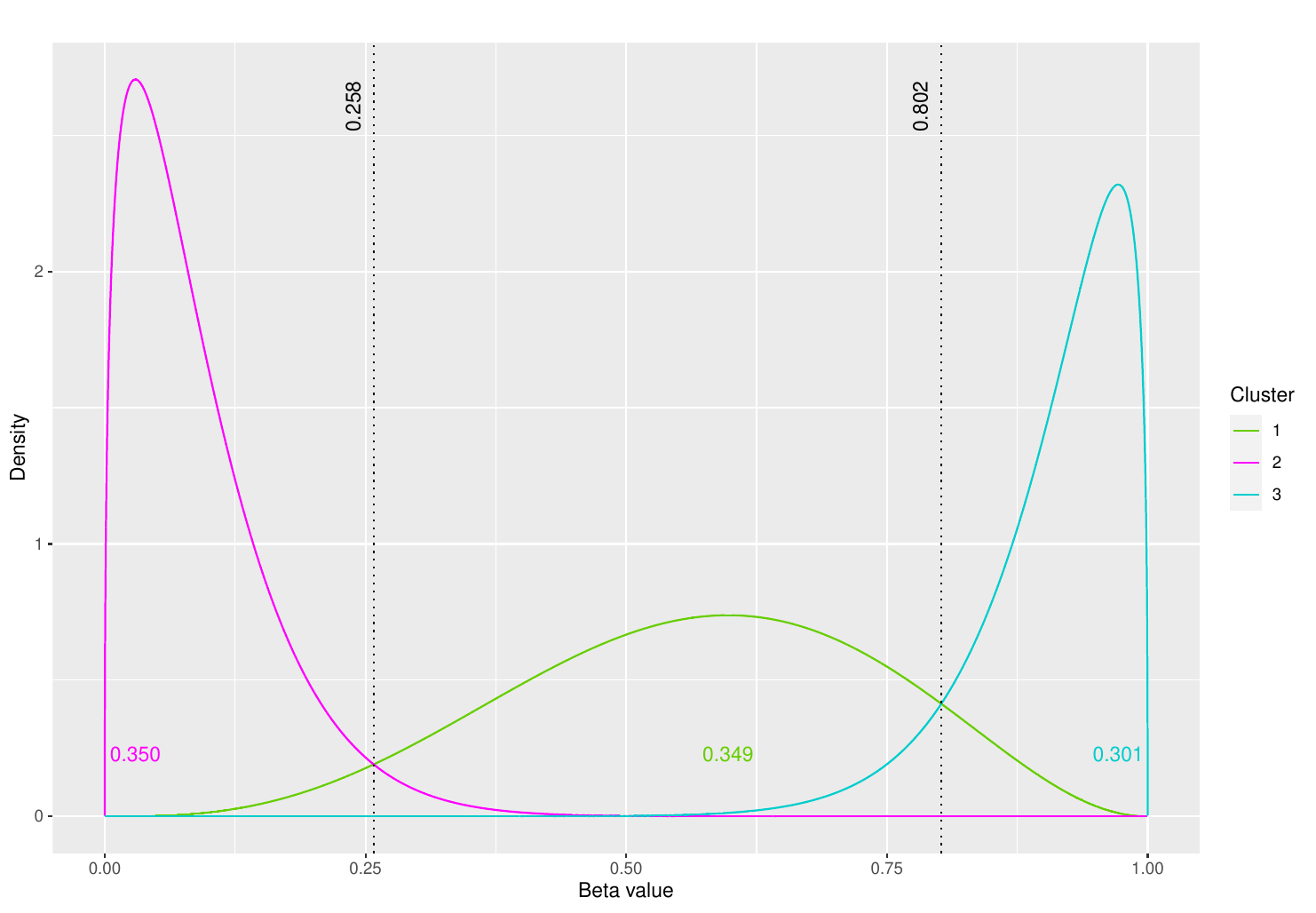}
\caption{\textbf{Fitted density estimates under the K$\cdot\cdot$ model on a simulated dataset from sample type A.}}
\caption*{\small{The thresholds between methylation states are illustrated by black dotted lines.
The estimated mixing proportions are also displayed.}}
\label{Fig3}
\end{center}
\end{figure}

To cluster the CpG sites in a sample type into 3 clusters representing the 3 methylation states and to infer the thresholds between these states, the K$\cdot\cdot$ and KN$\cdot$ models were fitted to the data from sample type A in each of the 100 simulated datasets. The true K$\cdot\cdot$ generating model was selected by AIC, BIC and ICL to be optimal in each case. Fig \ref{Fig3} illustrates the density estimates under the clustering solution of the K$\cdot\cdot$ model for a single simulated dataset. The hemimethylated CpG sites are clustered in cluster 1, while the hypomethylated and hypermethylated CpG sites are in clusters 2 and 3 respectively. The estimated mixing proportion of CpG sites for each cluster (see Fig \ref{Fig3}) are notably similar to the true mixing proportions. As parameters are constrained to be equal for each patient in the K$\cdot\cdot$ model, a single set of thresholds is inferred for all 4 patients. The threshold (see Fig \ref{Fig3}) of 0.258 indicates that any CpG site with a lower \textit{beta} value is likely to be hypomethylated. Similarly any CpG site with a \textit{beta} value greater than the second threshold of 0.802 is likely to be hypermethylated. These objectively inferred thresholds are very close to the true thresholds of 0.244 and 0.808.

The adjusted Rand index (ARI) \citep{ARI} gives a measure of agreement between two clustering solutions, where an ARI of 1 indicates full agreement.  The mean ARI across the 100 simulated datasets for the K$\cdot\cdot$ model was 0.9949 (s.d. 0.0002) and for the  KN$\cdot$ model was 0.9949 (s.d. 0.0002), demonstrating accurate and stable clustering solutions. The mean ARI when comparing the K$\cdot\cdot$ and KN$\cdot$ clustering solutions was 0.999 (s.d. 0.00001).  A summary of the parameter estimates and kernel density plots under the K$\cdot\cdot$ model are available in Appendices S4--S5.

\subsubsection*{\textbf{Identifying DMCs}}
To identify differentially methylated CpG sites between multiple DNA sample types in the simulated data, the K$\cdot$R model is fitted to each of the $C \times NR$ dimensional datasets. For each CpG site, as there are $R = 2$ sample types, $G^R = 9$ different combinations of the three methylation states are possible across sample types A and B. The CpG sites that are e.g., hypomethylated in one sample type and hypermethylated in the other are of interest as they indicate potential epigenetic changes in the genome.

 Under the K$\cdot$R model, for each simulated data set, the AIC, BIC and ICL criteria were non-informative as they consistently decreased for $K = 2, \ldots, 30$ (see Appendix S6). 
 Thus, the biologically motivated K$\cdot$R model with $K = G^R = 9$ was considered here. The AUC and WD metrics were employed to assess the similarity between the $R = 2$ probability distributions within each cluster, giving insight to the degree of differential methylation within clusters. Table \ref{Table 2a}  shows the mean and standard deviation of the AUC and WD values for each cluster across the 100 simulated datasets. Throughout, clusters are presented in descending order of their degree of differential methylation, based on decreasing AUC and, in the case of ties, WD values. 
The six differentially methylated clusters i.e., those in which the methylation state was different between the two sample types, are correctly highlighted as the most differentially methylated.

\begin{figure}
\begin{center}
\includegraphics[width=1\textwidth, height=13cm]{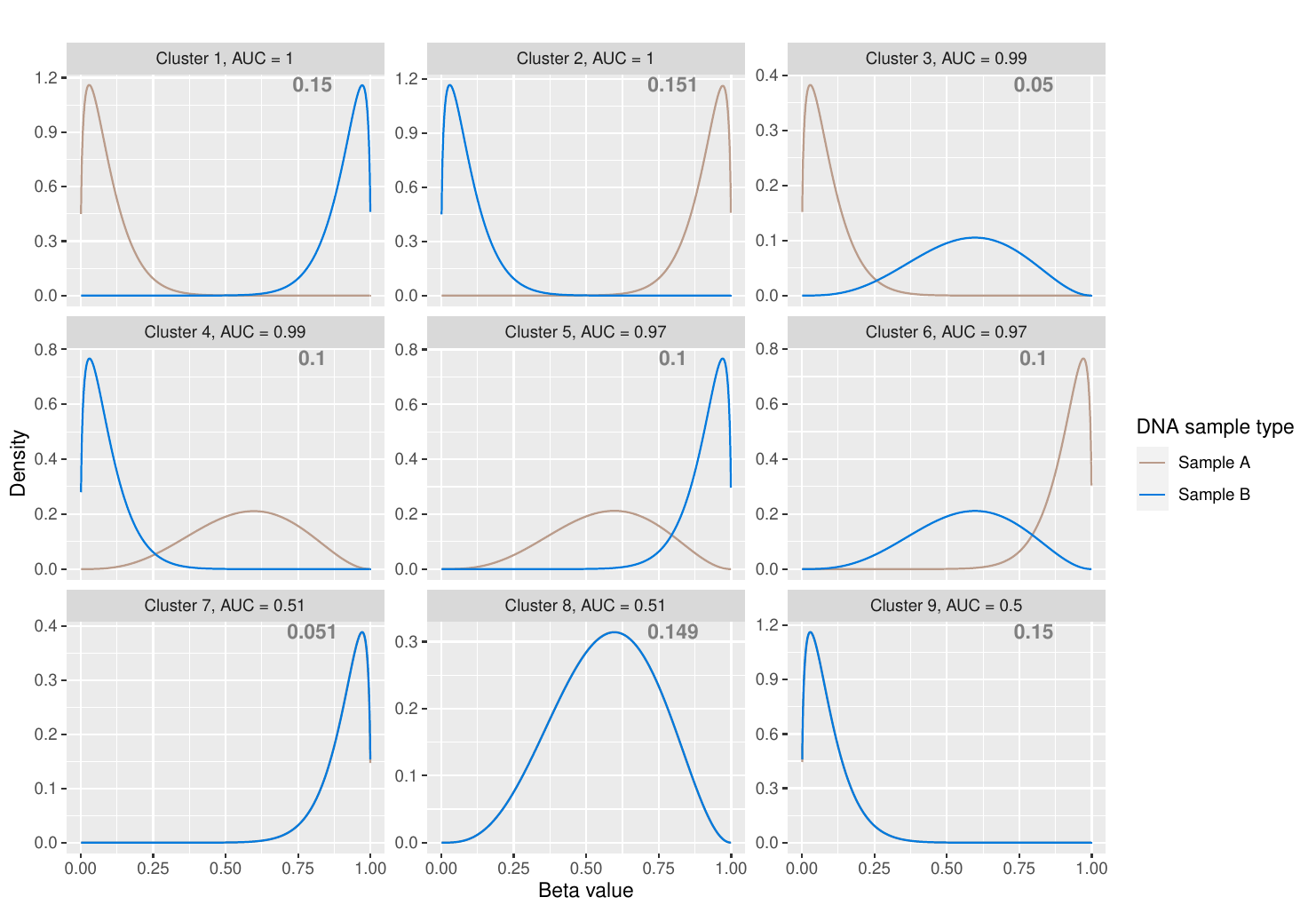}
\caption{\textbf{Fitted density estimates under the K$\cdot$R model on a simulated dataset.}}
\caption*{\small{The estimated mixing proportions are displayed in the relevant panel.}}
\label{Fig4}
\end{center}
\end{figure}

The graph in Fig \ref{Fig4} shows the fitted density estimates of the clustering solution under the K$\cdot$R model for a single simulated dataset.  The associated dissimilarity metrics correctly indicate clusters 1--6 as the most differentially methylated clusters, with 65.1\% of CpG sites belonging to these clusters, which is very close to the true mixing proportions.  The density estimates show that, for example, cluster 1 captures DMCs which are hypomethylated in sample type A and hypermethylated in sample type B while cluster 2 contains DMCs which are hypermethylated in sample type A and hypomethylated in sample type B. Mean standard performance metrics across the 100 simulated datasets signify an accurate and stable clustering process, with a mean false discovery rate (FDR) of 0.0041 (s.d. 0.0121), mean sensitivity of 0.9742 (s.d. 0.0563), mean specificity of 0.9921 (s.d. 0.0244) and mean ARI of 0.9758 (s.d. 0.0370). A summary of parameter estimates under the K$\cdot$R model is available in Appendix S4 with kernel density estimates in Appendix S7.

All computations were conducted using R \citep{R2024} on a Windows 11 operating system equipped with an Intel Core i7 CPU with 2.70GHz speed and 16GB RAM. In terms of computational cost, for example, fitting the K$\cdot$R model to a single simulated dataset took 55.86 seconds on a computer with 6 cores.
To explore the impact of an increasing value of $N$ on the computational cost, further simulation studies in which $N = \{8, \ldots, 60\}$, where $N$ increased in increments of 4, demonstrated a linear increase in computational cost. This is intuitive given the form of the  model's likelihood function with respect to $N$.
Further details on computational cost are provided, for all three BMM models, in Appendix S8.

The ability of the K$\cdot$R model to detect DMCs was compared with that of the state-of-the-art \texttt{limma} method \citep{limma}, which requires the beta values to be transformed into 
\textit{M}-values for analysis. 
Results indicate \texttt{limma} also performs well but with lower accuracy than the BMM approach: \texttt{limma} had a mean FDR of 0.0080 (s.d. 0.0001), mean sensitivity of 0.9074 (s.d. 0.0008), mean specificity of 0.9864 (s.d. 0.0003) and mean ARI of 0.7562 (s.d. 0.0018).  
The lower mean sensitivity value in particular suggests that \texttt{limma} identified fewer DMCs than the BMM. Boxplots displaying the performance metrics across the 100 simulations for the K$\cdot$R model and \texttt{limma} are available in Appendix S9.

Finally, to explore the robustness of the BMM approach to model misspecification, the same simulation settings were considered but where the data were simulated from a $t$-distribution with 8 degrees of freedom. Both the K$\cdot$R model and \texttt{limma} demonstrated mixed ability to detect DMCs when applied to the expit-transformed and logit-transformed data respectively. The BMM and \texttt{limma} approaches attained, respectively, a mean FDR of 0.4918 (s.d. 0.076) and 0.7018 (s.d. 0.0006), a mean sensitivity of 0.4558 (s.d. 0.0908) and 0.9995 (s.d. 0.0002) and a mean specificity of 0.875 (s.d. 0.0203) and 0.3279 (s.d. 0.0019). The BMM's low sensitivity and high specificity suggests that the BMM identified the non-DMCs correctly but failed to detect a large portion of true DMCs, while \texttt{limma} demonstrated the opposite ability. Boxplots of the performance metrics are provided in Appendix S10.

 \begin{table}
 \centering
 \caption{\textbf{Mean and standard deviation (s.d.) of the AUC and WD metrics for each cluster across the 100 simulated datasets. }}
\label{Table 2a}
 \begin{tabular}{l | c|c | c |c|c|c|c|c|c|c|} 
\cline{2-11}
 \multicolumn{1}{l|}{} & \multicolumn{10}{|c|}{Cluster} \\
 \cline{2-11}
 \multicolumn{2}{c|}{ } & 1 & 2 & 3 & 4 & 5 & 6 & 7 & 8 & 9 \\ 
 \cline{2-11}
  AUC& Mean & 1.0000 &  1.0000 &  0.9951 &  0.9940 &  0.9726 &  0.9456  & 0.5225 &  0.5139 &  0.5041 \\ 
 \cline{2-11}
   & S.D. & 0.0000 &  0.0000 &  0.0009 &  0.0009 &  0.0028 &  0.0508 &  0.0147 &  0.0115 &  0.0076 \\ 
 \cline{2-11}
 WD & Mean   & 0.8187  & 0.8185 &  0.4796 &  0.4797 &  0.3377 &  0.3162 &  0.0043 &  0.0040 &  0.0005 \\
 \cline{2-11}
  & S.D.  & 0.0001 &  0.0001 &  0.0005 &  0.0004 &  0.0012 &  0.0456 &  0.0091 &  0.0089 &  0.0015  \\
 \cline{2-11}
\end{tabular}
\end{table}

\subsection*{\textbf{Prostate cancer data results}}

\subsubsection*{\textbf{Estimating methylation state thresholds}}

For the PCa data, to cluster the CpG sites into the 3 methylation states and objectively infer the methylation state thresholds in the benign and tumour sample  types, the K$\cdot\cdot$ and KN$\cdot$ models were fitted. The AIC, BIC and ICL suggest the KN$\cdot$ model as optimal for both the benign and tumour sample types. This is intuitive, particularly for the tumour sample  types where the degree of disease varies for each patient, as the KN$\cdot$ model allows for patient specific shape parameters. 

The fitted density estimates and inferred thresholds for patient 1 are discussed here; those for patients 2, 3 and 4 are available in Appendix S11. 

In the benign sample type from patient 1, the estimated mixing proportions were 0.244 for hypomethylation, 0.363 for hemimethylation, and 0.393 for hypermethylation. The inferred methylation state thresholds are 0.258 and 0.747 for the benign sample type, and 0.19 and 0.751 for the tumour sample type. The hypermethylation state thresholds in the benign and tumour sample types are very close; in contrast, the hypomethylation state thresholds are quite different. While these objective thresholds are close to the subjective values suggested in the literature of 0.2 and 0.8, the difference results in more hypo- and hypermethylated CpG sites being identified by the BMM as DMCs.

Patient 1 was known to have a greater degree of disease severity than the other patients with a matched normal DNA methylation profile more tumour-like than benign. For patient 1, the hypermethylation threshold (0.747) was lower than that for the other patients (0.774, 0.766 and 0.814 for patients 2, 3 and 4 respectively) suggesting more hypermethylated CpG sites in patient 1's benign sample type than in the other patients' samples. A similar pattern was observed in the methylation thresholds inferred from the patients' tumour sample types (see Appendix S11) in that the threshold was lower for patient 1.

The ARIs of 0.94 and 0.96 between the KN$\cdot$ and K$\cdot\cdot$ solutions for the benign and tumour sample types respectively indicate good clustering agreement. Summaries of parameter estimates and the kernel density estimates under the KN$\cdot$ model are available in Appendices S4 and S12.

\subsubsection*{\textbf{Identifying DMCs in the PCa data}}
To identify differentially methylated CpG sites in the PCa data, the K$\cdot$R model was fitted to the $C \times NR$ dimensional dataset. Similar to the simulation study, the AIC, BIC and ICL were non-informative and consistently decreased across models with $K = 2, \ldots, 30$ (see Appendix S13). 
Thus, motivated by the $G^R = 9$ unique methylation state combinations that could be biologically present across the benign and tumor sample types, a model with $K = G^R = 9$ was fitted to the PCa data. 

\begin{figure}
\begin{center}
\includegraphics[width=1\textwidth, height =15cm]{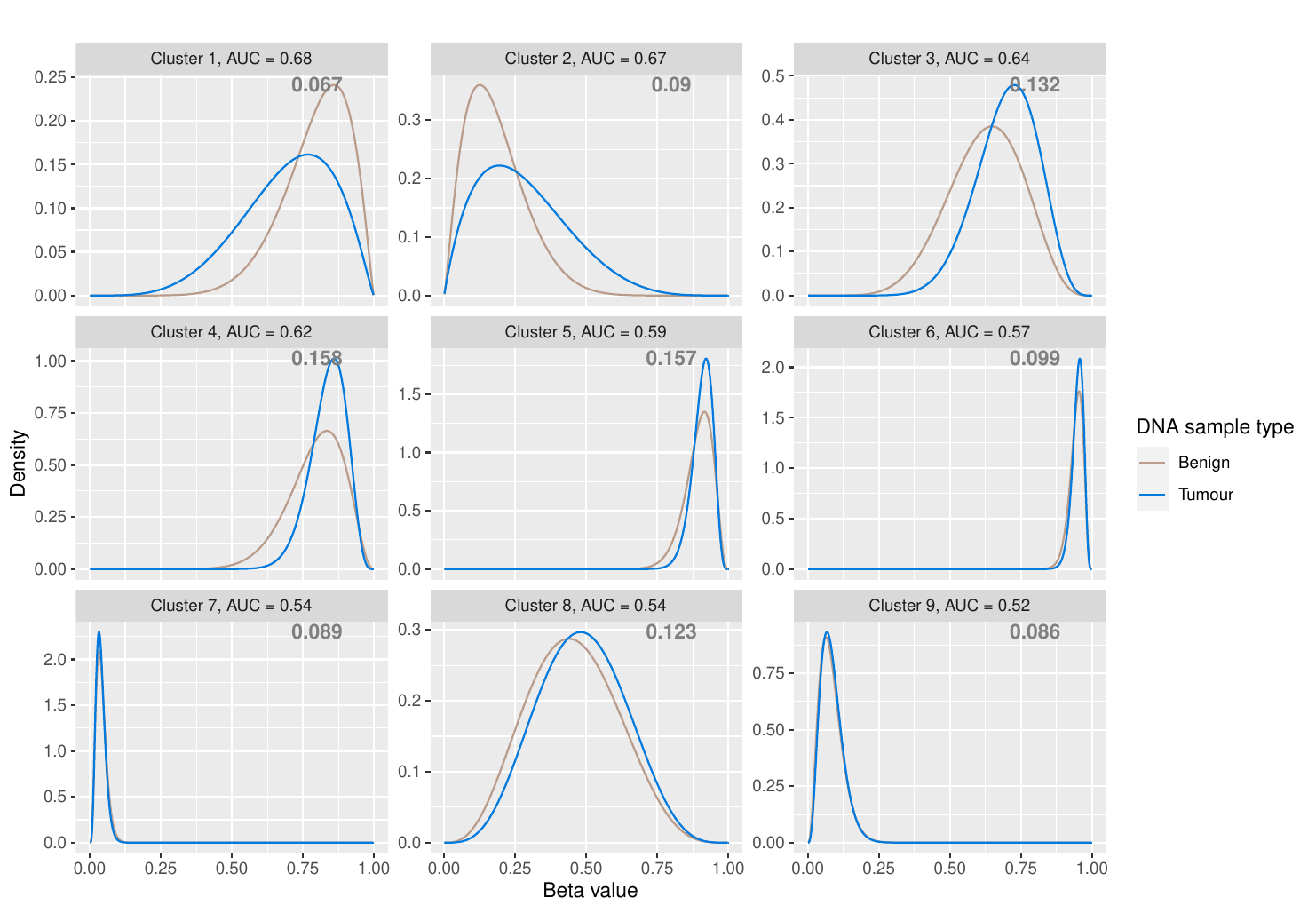}
\caption{\textbf{Fitted density estimates under the clustering solution of the K$\cdot$R model.}}
\caption*{\small{The model estimates parameters for $K = 9$ clusters for the DNA methylation data from benign and tumour prostate sample types. The estimated mixing proportions are displayed in the relevant panel.}}
\label{Fig5}
\end{center}
\end{figure}

\begin{figure}
\begin{center}
\includegraphics[width=0.8\textwidth, height =9cm]{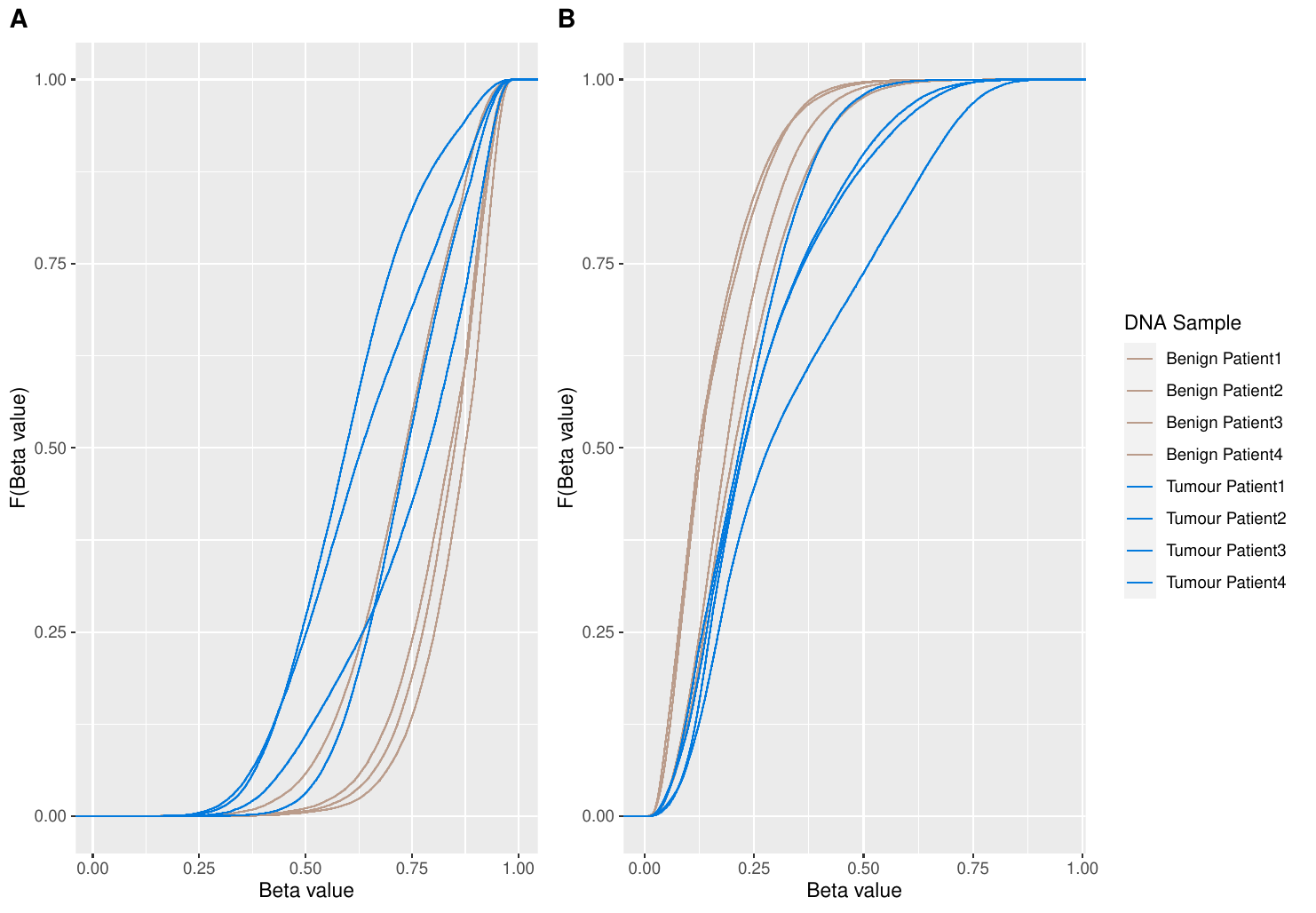}
\caption{\textbf{Empirical cumulative distribution functions for DMCs.}} 
\caption*{\small{Empirical cumulative distribution functions for DMCs in (A) cluster 1 and (B) cluster 2 for all patients and sample types.}}
\label{Fig6}
\end{center}
\end{figure}

Fig \ref{Fig5} illustrates fitted density estimates of the clustering solution, ordered by AUC or WD in the case of equal AUC values. Kernel density estimates are in Appendix S14.    
Table \ref{Table 4} summarises the parameter estimates and details the AUC and WD metrics for each cluster. 
As the extent of disease progression varied across patients, the PCa data were noisier than the simulated data, and the AUC and WD metrics were lower in general. 
Both the AUC and WD  metrics suggest that clusters 1 and 2 contain the CpG sites that are most differentially methylated in nature. Inspection of the density and parameter estimates of clusters 1 and 2 provides insight: cluster 1 captures CpG sites exhibiting a downward trend in methylation values for tumor samples with increased methylation values in the benign samples. On the other hand, CpG sites in cluster 2 tend to have higher methylation levels in tumor samples than in benign samples. 
While there are visual differences between the benign and tumor density estimates in clusters 3--5, the density estimates in later clusters are almost visually indistinguishable between the two samples, particularly in the case of clusters 6, 7 and 9. This is intuitive as clusters with smaller AUC (and WD) values contain the least differentially methylated CpG sites between the benign and tumour samples and their respective density estimates within such clusters will be very similar. 
Fig \ref{Fig6} shows the empirical cumulative distribution functions (ECDFs) for the DMCs within clusters 1 and 2, for both benign and tumor sample types. The ECDF for cluster 1 shows an increase in \textit{beta} values within the benign samples, relative to the tumor samples while the ECDF of the CpG sites in cluster 2 indicates an elevation in \textit{beta} values in the tumor samples, compared to the benign samples. The K$\cdot$R model identifies 102,757 CpG sites, belonging to clusters 1 and 2, as being mostly differentially methylated. 

\begin{table}
    \caption{\textbf{Beta distributions' parameter estimates for the benign and tumour samples, and the AUC and WD metrics, for the PCa data under the K$\cdot$R model. }}
    \label{Table 4}
      \centering
 \begin{tabular}{|c | c | c | c| c|c |  c | c| c|c|c|} 
  \hline
& \multicolumn{4}{c|}{ Benign} &\multicolumn{4}{c|}{ Tumour}& & \\ 
\hline
 Cluster & $\hat{\alpha}$  & $\hat{\delta}$    & Mean & S.D.& $\hat{\alpha}$  & $\hat{\delta}$    & Mean & S.D. & AUC & WD \\ 
 \hline
   1 &8.815     & 2.277    &  0.795 &0.116  & 5.076 & 2.231 & 0.695 & 0.160 &0.683 & 0.100 \\ 
 \hline
   2 &2.324     & 10.223    &0.185   & 0.106& 1.975 & 5.040 &0.282  & 0.159& 0.667 & 0.096  \\
 \hline
   3 & 8.005   & 4.810    & 0.625 & 0.130& 12.058 &5.170  &  0.700& 0.107 &0.639 & 0.075 \\
 \hline
    4 &13.006    & 3.387    &0.793   &0.097  & 27.000 & 5.249 & 0.837 &  0.064 &0.624 & 0.044 \\ 
 \hline
   5 & 33.720   & 4.006    &0.894   &  0.050 & 56.506 & 5.734 & 0.908 & 0.036 &0.593&0.015\\
 \hline
   6 & 84.926  & 5.023    & 0.944  &  0.024& 111.727 & 5.978 & 0.949 & 0.020 &0.572&0.005 \\ 
 \hline
    7 & 4.842    & 112.897  & 0.041  & 0.018 &  5.455& 133.043 &0.039  &  0.016&0.542&0.002 \\
 \hline
   8 &4.071     &4.924     & 0.453  & 0.157  & 4.686 & 4.990 & 0.484 & 0.153 &0.537 &0.032\\
 \hline
   9 & 3.749    & 41.317   & 0.083  &  0.041& 4.194 & 45.197 &0.085  & 0.039 &0.523&0.002\\
 \hline
        \end{tabular}

        \footnotesize{$\ast$ Standard deviation is denoted as S.D.}

\end{table}

On performing gene ontology analysis \citep{gometh}, CpG sites in cluster 2 were found to be related to known genes e.g., RARB, GSTP1, RASSF1, SFRP2, which are implicated in prostate cancer. For example, hypermethylation of RARB promoter genes is a significant biomarker in diagnosing prostate cancer \citep{RARB}.
The methylation levels of the DMCs in cluster 2 that belong to the RARB genes suggest the median \textit{beta} value is higher in the tumour sample type than in the benign sample type for all patients (see Appendix S15). 
Through non-parametric tests, the \textit{beta} values were shown to be significantly higher in the tumour samples than in the benign samples for the CpG sites related to these genes ($p < 0.05$). Further, the ECDF for DMCs related to the RARB genes for benign and tumour sample types illustrated that the DMCs have increased \textit{beta} values in the tumour samples compared to the benign samples (see Appendix S16). Analysis of genes linked to CpG sites in clusters 3--9, the less differentially methylated clusters, did identify some genes previously implicated in prostate cancer e.g., AKT1 \citep{AKT1}. However, non-parametric tests also suggested no statistically significant difference ($p > 0.05$) in \textit{beta} values between benign and tumour samples for CpG sites in clusters 3--9 linked to the gene AKT1, as did associated box and  ECDF plots (see Appendix S17).

Gene ontology analysis of the DMCs in cluster 1 also unveiled approximately 16 noteworthy biological processes. These processes, distinct from cancer-related pathways, encompass vital functions such as nervous system processes.
A substantial count of 1001 significant biological processes were revealed among the DMCs in cluster 2 (FDR-adjusted $p$-value $< 0.05$). Further, considering the KEGG pathways, the DMCs in cluster 1 were associated with one significant pathway, while the DMCs within cluster 2 exhibited involvement in a noteworthy 61 significant pathways.
Of these significant pathways, many were cancer related e.g., the proteoglycans in cancer pathway was the second most enriched pathway.

Given the BMM's model-based approach to clustering, the uncertainty in CpG site $c$'s clustering is available as $1 - \max\limits_{k = 1, \ldots, K}(\hat{z}_{ck})$ , with a maximum possible uncertainty of $1-1/K = 8/9$. 
All CpG sites have clustering uncertainties well below this maximum, demonstrating that the CpG sites are clustered with high certainty (see Appendix S18).

To demonstrate the general applicability of the approach, the BMMs were also fitted to a publicly available DNA methylation dataset from an esophageal squamous cell carcinoma study (ESCC); full details are available in Appendix S19.

\section*{Discussion}
DNA methylation is widely studied for disease diagnosis and treatment. Technology advancements have led to the development of microarrays that can assay e.g., 850,000 CpG sites from a DNA sample 
\citep{Ruth}, but the analysis of these large arrays has been limited by a lack of appropriate statistical methods for the bounded and heteroskedastic nature of \textit{beta}-valued DNA methylation data. The methylation states of CpG sites are often of interest and are typically identified using thresholds which are defined in the literature based on intuition \citep{ChenXin} rather than using an objective approach. Additionally, to detect DMCs, it is common practice to apply a logit transformation to \textit{beta} values, and subsequently model them as Gaussian-distributed \citep{Du, limma}. 
Alternatively, comparisons between untransformed methylation levels among sample types are often conducted using multiple moderated t-tests or Wilcoxon rank sum tests \citep{IMA,COHCAP}. The approach proposed here advocates against transforming the data and instead proposes modelling the data in its innate form when inferring methylation state thresholds and DMCs.

In the context of prostate cancer, a family of beta mixture models is proposed which employs novel constraints on the model parameters to cluster CpG sites based on untransformed \textit{beta} values
to objectively identify methylation state thresholds and
DMCs between benign and tumour samples.
The BMMs use a model-based clustering approach and inference is computationally efficient through the use of a digamma approximation.
The objective inference of methylation thresholds demonstrated that the thresholds of 0.2 and 0.8 or 0.3 and 0.7 defined in literature are not appropriate for every scenario. The thresholds inferred from each patient's data showed variability, reflecting the different stages of disease among patients. The proposed K$\cdot$R model clusters CpG sites from multiple DNA sample types to determine the CpG sites with differential methylation. Gene ontology enrichment analysis of the genes associated with CpG sites in the most differentially methylated clusters revealed several significant biological processes, cancer-related pathways and genes implicated in prostate cancer, opening new avenues of research. The results illustrate the ability of the BMMs to analyse large microarrays consisting of samples from multiple conditions from several patients and to reveal biologically relevant methylation patterns, thus contributing to advances in the field of quantitative DNA methylation analysis.

In terms of the family of BMMs developed here, there are several potential future research directions. For example, while DNA methylation can be influenced by  environmental and clinical variables, the proposed BMMs do not incorporate such covariates. However, the BMMs could be extended, for example using a mixture of experts approach \citep{Claire} where the parameters of the BMM are modeled as functions of the covariates, to offer a richer modelling framework.
Further, a key assumption of the proposed BMMs is that the methylation states of adjacent CpG sites are conditionally independent given their cluster membership. However, methylation levels of adjacent CpG sites are often highly correlated
\citep{Hodges}.  This phenomenon gives rise to the emergence of biologically meaningful regions with discernible patterns. Expanding the scope of the BMM family to encompass the spatial dependencies within the data would present an opportunity to incorporate these structural nuances and ultimately facilitate the identification of particularly relevant differentially methylated regions. While the scale of missing data in the prostate cancer data considered here was almost negligible, it could be more prevalent in other settings. Such cases would motivate the development of imputation approaches that are cognisant of the heterogeneity typical of DNA methylation datasets. Finally, the methylation state of a human genome changes over time depending on clinical conditions. Longitudinal methylation data are often collected to study the effect of environmental changes or treatments on disease progression. Such data are vast and current approaches struggle to handle these extensive data in their innate form. In order to analyze methylation changes over time in multiple patients, similar to \cite{Gift}, the BMMs could be further enhanced to model dependency over time.



\section{\textbf{Acknowledgements}}
This publication has emanated from research conducted with the financial support of Science Foundation Ireland under Grant number 18/CRT/6049. For the purpose of Open Access, the author has applied a CC BY public copyright licence to any Author Accepted Manuscript version arising from this submission The authors wish to thank members of the Working Group in Model-based Clustering for their discussions on this work.

 \bibliographystyle{authordate1}

\bibliography{Main.bib}{}   

\appendix

\section{Supporting Information for \lq A novel family of beta mixture models for the differential analysis of DNA methylation data: an application to prostate cancer\rq  \space data by Majumdar et al.}

\subsection*{Appendix S1}
\paragraph{{\textbf{K$\cdot\cdot$ Model}}} 
The complete data log-likelihood for this model is,
\begin{equation*}
\begin{aligned}
  \ell_{C}(\boldsymbol{\tau},\boldsymbol{\theta},\mathbf{Z}|\mathbf{X}) = \sum\limits_{c=1}^{C}\sum\limits_{k=1}^{K} z_{ck} \{\log  \tau_k + 
  \sum\limits_{n=1}^{N} \sum\limits_{r=1}^{1}
   \log [\mbox{Beta}(x_{cnr};\alpha_{k\cdot\cdot},\delta_{k\cdot\cdot})]\}.
   \end{aligned}
\end{equation*}

In the Expectation-step of the EM algorithm the $\hat{z}_{ck}$ is calculated given the current parameter estimates. In the Maximisation-step the expected complete data log-likelihood function to be optimized is,
\begin{equation}
\label{equation 1}
\begin{aligned}
    \ell_{C}(\boldsymbol{\tau,\theta}|\mathbf{X},\mathbf{\hat{Z}}) =  &\sum\limits_{c=1}^C \sum\limits_{k=1}^K \hat{z}_{ck} \{\log \tau_k + \\
    & \sum\limits_{n=1}^N\sum\limits_{r=1}^{1}[(\alpha_{k\cdot\cdot}-1)\log x_{cnr} + 
    (\delta_{k\cdot\cdot}-1)\log (1-x_{cnr})-\log \mbox{B}(\alpha_{k\cdot\cdot},\delta_{k\cdot\cdot})]\}.
    \end{aligned}
\end{equation}
Differentiating (\ref{equation 1}) w.r.t $\alpha_{k\cdot\cdot}$ yields,
\begin{equation}
\label{equation 2}
\begin{aligned}
    \frac{\partial{\ell_{C}}}{\partial{\alpha_{k\cdot\cdot}}} =  \sum\limits_{c=1}^C \hat{z}_{ck}  \{\log x_{cnr} - 
    [\psi(\alpha_{k\cdot\cdot})-\psi(\alpha_{k\cdot\cdot}+\delta_{k\cdot\cdot})]\}
   \end{aligned}
\end{equation}
where $\psi$ is the digamma function.

Similarly, the derivative of $\ell_{C}(\boldsymbol{\tau}, \boldsymbol{\theta}|\mathbf{X},\mathbf{\hat{Z}})$ w.r.t $\delta_{k\cdot\cdot}$ is,
\begin{equation}
\label{equation 3}
\begin{aligned}
    \frac{\partial{\ell_{C}}}{\partial{\delta_{k\cdot\cdot}}} =  \sum\limits_{c=1}^C \hat{z}_{ck} \{\log (1-x_{cnr}) -  [\psi(\delta_{k\cdot\cdot})-\psi(\alpha_{k\cdot\cdot}+\delta_{k\cdot\cdot})]\}.
    \end{aligned}
\end{equation}
The lower bound value of the digamma function ($ \psi(y) > \log(y-1/2)$) is used in (\ref{equation 2}) and (\ref{equation 3}) to get closed-form
solutions at the Maximisation-step of the EM algorithm,
\begin{equation}
\label{equation 4}
\begin{aligned}
    \frac{\partial{\ell_{C}}}{\partial{\alpha_{k\cdot\cdot}}} \approx  \sum\limits_{c=1}^C \hat{z}_{ck} \sum\limits_{n=1}^N \sum\limits_{r=1}^{1} \left[ \log x_{cnr} - 
    \log \frac{\alpha_{k\cdot\cdot}-1/2}{\alpha_{k\cdot\cdot}+\delta_{k\cdot\cdot}-1/2} \right]
   \end{aligned}
\end{equation}
and
\begin{equation}
\label{equation 5}
\begin{aligned}
    \frac{\partial{\ell_{C}}}{\partial{\delta_{k\cdot\cdot}}} \approx  \sum\limits_{c=1}^C \hat{z}_{ck} \sum\limits_{n=1}^N \sum\limits_{r=1}^{1}\left[ \log (1-x_{cnr}) -  
    \log \frac{\delta_{k\cdot\cdot}-1/2}{\alpha_{k\cdot\cdot}+\delta_{k\cdot\cdot}-1/2} \right].
    \end{aligned}
\end{equation}

Equating (\ref{equation 4}) and (\ref{equation 5}) to zero, we get the approximate estimates of $\alpha_{k\cdot\cdot}$ and $\delta_{k\cdot\cdot}$ as,
\begin{equation*}
    \begin{aligned}
        \alpha_{k\cdot\cdot} = 0.5+ \frac{0.5\exp (-y_2)}{\{[\exp(-y_2)-1][\exp(-y_1)-1]\}-1}
    \end{aligned}
\end{equation*}
and
\begin{equation*}
    \begin{aligned}
        \delta_{k\cdot\cdot} =  \frac{0.5\exp (-y_2)[\exp(-y_1)-1]}{\{[\exp(-y_2)-1][\exp(-y_1)-1]\}-1},
    \end{aligned}
\end{equation*}
where
        $y_1 = (\sum\limits_{c=1}^C z_{ck} \log x_{cnr})/(N \sum\limits_{c=1}^C z_{ck})$
and 
        $y_2 = (\sum\limits_{c=1}^C z_{ck} \log (1-x_{cnr}))/(N \sum\limits_{c=1}^C z_{ck})$.

\clearpage

\section*{Appendix S2}
\paragraph{{\textbf{KN$\cdot$ Model}}} 
The complete data log-likelihood for this model is,
\begin{equation*}
\begin{aligned}
  \ell_{C}(\boldsymbol{\tau},\boldsymbol{\theta},\mathbf{Z}|\mathbf{X}) = \sum\limits_{c=1}^{C}\sum\limits_{k=1}^{K} z_{ck} \{\log  \tau_k + 
  \sum\limits_{n=1}^{N} \sum\limits_{r=1}^{1}
   \log [\mbox{Beta}(x_{cnr};\alpha_{kn\cdot},\delta_{kn\cdot})]\}.
   \end{aligned}
\end{equation*}

In the Expectation-step of the EM algorithm the $\hat{z}_{ck}$ is calculated given the current parameter estimates. In the Maximisation-step the expected complete data log-likelihood function to be optimized is,
\begin{equation}
\label{equation 6}
\begin{aligned}
    \ell_{C}(\boldsymbol{\tau,\theta}|\mathbf{X},\mathbf{\hat{Z}}) = & \sum\limits_{c=1}^C \sum\limits_{k=1}^K \hat{z}_{ck} \{\log \tau_k + \\
    &\sum\limits_{n=1}^N\sum\limits_{r=1}^{1}[(\alpha_{kn\cdot}-1)\log x_{cnr} + 
    (\delta_{kn\cdot}-1)\log (1-x_{cnr})-\log \mbox{B}(\alpha_{kn\cdot},\delta_{kn\cdot})]\}.
    \end{aligned}
\end{equation}
Differentiating (\ref{equation 6}) w.r.t $\alpha_{kn\cdot}$ yields,
\begin{equation}
\label{equation 7}
\begin{aligned}
    \frac{\partial{\ell_{C}}}{\partial{\alpha_{kn\cdot}}} =  \sum\limits_{c=1}^C \hat{z}_{ck} \{\log x_{cnr} - 
    [\psi(\alpha_{kn\cdot})-\psi(\alpha_{kn\cdot}+\delta_{kn\cdot})]\}
   \end{aligned}
\end{equation}
where $\psi$ is the digamma function.

Similarly, the derivative of $\ell_{C}(\boldsymbol{\tau}, \boldsymbol{\theta}|\mathbf{X},\mathbf{\hat{Z}})$ w.r.t $\delta_{kn\cdot}$ is,
\begin{equation}
\label{equation 8}
\begin{aligned}
    \frac{\partial{\ell_{C}}}{\partial{\delta_{kn\cdot}}} =  \sum\limits_{c=1}^C \hat{z}_{ck} \{\log (1-x_{cnr}) -  [\psi(\delta_{kn\cdot})-\psi(\alpha_{kn\cdot}+\delta_{kn\cdot})]\}.
    \end{aligned}
\end{equation}
The lower bound value of the digamma function ($ \psi(y) > \log(y-1/2)$)  is used in (\ref{equation 7}) and (\ref{equation 8}) to get closed-form
solutions at the Maximisation-step of the EM algorithm,
\begin{equation}
\label{equation 9}
\begin{aligned}
    \frac{\partial{\ell_{C}}}{\partial{\alpha_{kn\cdot}}} \approx  \sum\limits_{c=1}^C \hat{z}_{ck} \sum\limits_{n=1}^N \sum\limits_{r=1}^{1} \left[ \log x_{cnr} - 
    \log \frac{\alpha_{kn\cdot}-1/2}{\alpha_{kn\cdot}+\delta_{kn\cdot}-1/2} \right]
   \end{aligned}
\end{equation}
and
\begin{equation}
\label{equation 10}
\begin{aligned}
    \frac{\partial{\ell_{C}}}{\partial{\delta_{kn\cdot}}} \approx  \sum\limits_{c=1}^C \hat{z}_{ck} \sum\limits_{n=1}^N \sum\limits_{r=1}^{1}\left[ \log (1-x_{cnr}) -  
    \log \frac{\delta_{kn\cdot}-1/2}{\alpha_{kn\cdot}+\delta_{kn\cdot}-1/2} \right].
    \end{aligned}
\end{equation}

Equating (\ref{equation 9}) and (\ref{equation 10}) to zero, we get the approximate estimates of $\alpha_{kn\cdot}$ and $\delta_{kn\cdot}$ as,
\begin{equation*}
    \begin{aligned}
        \alpha_{kn\cdot} = 0.5+ \frac{0.5\exp (-y_2)}{\{[\exp(-y_2)-1][\exp(-y_1)-1]\}-1}
    \end{aligned}
\end{equation*}
and
\begin{equation*}
    \begin{aligned}
        \delta_{kn\cdot} =  \frac{0.5\exp (-y_2)[\exp(-y_1)-1]}{\{[\exp(-y_2)-1][\exp(-y_1)-1]\}-1},
    \end{aligned}
\end{equation*}
where
        $y_1 = (\sum\limits_{c=1}^C z_{ck} \log x_{cnr})/( \sum\limits_{c=1}^C z_{ck})$
and 
        $y_2 = (\sum\limits_{c=1}^C z_{ck} \log (1-x_{cnr}))/( \sum\limits_{c=1}^C z_{ck})$.

\section*{Appendix S3}
\paragraph{{\textbf{K$\cdot$R Model}}} 
The complete data log-likelihood for this model is,
\begin{equation*}
\begin{aligned}
  \ell_{C}(\boldsymbol{\tau},\boldsymbol{\theta},\mathbf{Z}|\mathbf{X}) = \sum\limits_{c=1}^{C}\sum\limits_{k=1}^{K} z_{ck} \{\log  \tau_k + 
  \sum\limits_{n=1}^{N} \sum\limits_{r=1}^{R}
   \log [\mbox{Beta}(x_{cnr};\alpha_{k\cdot r},\delta_{k\cdot r})]\}.
   \end{aligned}
\end{equation*}

In the Expectation-step of the EM algorithm the $\hat{z}_{ck}$ is calculated given the current parameter estimates. In the Maximisation-step the expected complete data log-likelihood function to be optimized is,
\begin{equation}
\label{equation 11}
\begin{aligned}
    \ell_{C}(\boldsymbol{\tau,\theta}|\mathbf{X},\mathbf{\hat{Z}}) = & \sum\limits_{c=1}^C \sum\limits_{k=1}^K \hat{z}_{ck} \{ \log \tau_k + \\
    &\sum\limits_{n=1}^N\sum\limits_{r=1}^{R}[(\alpha_{k\cdot r}-1)\log x_{cnr} + 
    (\delta_{k\cdot r}-1)\log (1-x_{cnr})-\log \mbox{B}(\alpha_{k\cdot r},\delta_{k\cdot r})]\}.
    \end{aligned}
\end{equation}
Differentiating (\ref{equation 11}) w.r.t $\alpha_{k\cdot r}$ yields,
\begin{equation}
\label{equation 12}
\begin{aligned}
    \frac{\partial{\ell_{C}}}{\partial{\alpha_{k\cdot r}}} =  \sum\limits_{c=1}^C \hat{z}_{ck} \{\log x_{cnr} - 
    [\psi(\alpha_{k\cdot r})-\psi(\alpha_{k\cdot r}+\delta_{k\cdot r})]\}
   \end{aligned}
\end{equation}
where $\psi$ is the digamma function.

Similarly, the derivative of $\ell_{C}(\boldsymbol{\tau}, \boldsymbol{\theta}|\mathbf{X},\mathbf{\hat{Z}})$ w.r.t $\delta_{k\cdot r}$ is,
\begin{equation}
\label{equation 13}
\begin{aligned}
    \frac{\partial{\ell_{C}}}{\partial{\delta_{k\cdot r}}} =  \sum\limits_{c=1}^C \hat{z}_{ck} \{\log (1-x_{cnr}) -  [\psi(\delta_{k\cdot r})-\psi(\alpha_{k\cdot r}+\delta_{k\cdot r})]\}.
    \end{aligned}
\end{equation}
The lower bound value of the digamma function ($ \psi(y) > \log(y-1/2)$) is used in (\ref{equation 12}) and (\ref{equation 13}) to get closed-form
solutions at the Maximisation-step of the EM algorithm,
\begin{equation}
\label{equation 14}
\begin{aligned}
    \frac{\partial{\ell_{C}}}{\partial{\alpha_{k\cdot r}}} \approx  \sum\limits_{c=1}^C \hat{z}_{ck} \sum\limits_{n=1}^N \sum\limits_{r=1}^{1} \left[ \log x_{cnr} - 
    \log \frac{\alpha_{k\cdot r}-1/2}{\alpha_{k\cdot r}+\delta_{k\cdot r}-1/2} \right]
   \end{aligned}
\end{equation}
and
\begin{equation}
\label{equation 15}
\begin{aligned}
    \frac{\partial{\ell_{C}}}{\partial{\delta_{k\cdot r}}} \approx  \sum\limits_{c=1}^C \hat{z}_{ck} \sum\limits_{n=1}^N \sum\limits_{r=1}^{1}\left[ \log (1-x_{cnr}) -  
    \log \frac{\delta_{k\cdot r}-1/2}{\alpha_{k\cdot r}+\delta_{k\cdot r}-1/2} \right].
    \end{aligned}
\end{equation}

Equating (\ref{equation 14}) and (\ref{equation 15}) to zero, we get the approximate estimates of $\alpha_{knr}$ and $\delta_{knr}$ as,
\begin{equation*}
    \begin{aligned}
        \alpha_{k\cdot r} = 0.5+ \frac{0.5\exp (-y_2)}{\{[\exp(-y_2)-1][\exp(-y_1)-1]\}-1}
    \end{aligned}
\end{equation*}
and
\begin{equation*}
    \begin{aligned}
        \delta_{k\cdot r} =  \frac{0.5\exp (-y_2)[\exp(-y_1)-1]}{\{[\exp(-y_2)-1][\exp(-y_1)-1]\}-1},
    \end{aligned}
\end{equation*}
where
        $y_1 = (\sum\limits_{c=1}^C z_{ck} \log x_{cnr})/(N \sum\limits_{c=1}^C z_{ck})$
and 
        $y_2 = (\sum\limits_{c=1}^C z_{ck} \log (1-x_{cnr}))/(N \sum\limits_{c=1}^C z_{ck})$.

\clearpage
\subsection*{Appendix S4 }

\begin{table}[H]
 \centering
 \caption{Beta distributions' parameter estimates for sample type A in a simulated dataset under the K$\cdot\cdot$ model}
\label{Table 1}
 \begin{tabular}{c  c  c  c c} 
 \hline
  Clusters & $\hat{\alpha}$  & $\hat{\delta}$   & Mean & Std. deviation  \\ 
 \hline
   1 & 4.161 & 3.129 & 0.571&0.336   \\ 
 \hline
   2 & 1.396 & 14.092 & 0.090 &0.077  \\
 \hline
   3 &  13.761 &  1.371 & 0.909& 0.273 \\
 \hline
\end{tabular}
\end{table}

\begin{table}[!htb]
    \caption{Beta distributions' parameter estimates for sample type A in a simulated dataset under the K$\cdot$R model}. \label{Table 2}
    \setlength\tabcolsep{2.4pt}
    \begin{subtable}{.5\linewidth}
      \centering
        \caption{Sample A}
 \begin{tabular}{c  c  c  c c} 
 \hline
 Clusters & $\hat{\alpha}$  & $\hat{\delta}$   & Mean & Std. deviation  \\ 
 \hline
   1 & 1.391 & 14.024  &0.090& 0.077  \\ 
 \hline
   2 &  13.918 & 1.386   &0.909& 0.273 \\
 \hline
   3  & 1.384 & 13.954  &0.090& 0.078\\
 \hline
    4 &  4.168& 3.137  &0.571& 0.335  \\ 
 \hline
   5 & 4.157 &  3.134  &0.570& 0.335 \\
 \hline
   6 & 13.832 & 1.383  &0.909& 0.274 \\
 \hline
    7 &13.987 & 1.398  &0.909& 0.274  \\ 
 \hline
   8 & 4.155 &  3.134 &0.570& 0.335 \\
 \hline
   9 & 1.395 &  14.088 &0.090& 0.077 \\
 \hline
        \end{tabular}
        
    \end{subtable}
    \hspace*{0.5cm}
    \begin{subtable}{.5\linewidth}
      \centering
        \caption{Sample B}
 \begin{tabular}{c  c   c  c c} 
 \hline
Clusters & $\hat{\alpha}$  & $\hat{\delta}$   & Mean & Std. deviation  \\ 
 \hline
   1 & 13.908 & 1.383 & 0.091 &  0.273 \\ 
 \hline
   2 & 1.393  & 14.060 &  0.090& 0.077 \\
 \hline
   3 & 4.186 & 3.154 &  0.570&  0.335 \\
 \hline
    4 & 1.411 & 14.207 &  0.090& 0.077 \\ 
 \hline
   5 & 13.909 & 1.391 & 0.909 & 0.274 \\
 \hline
   6 & 4.156 & 3.128  &0.571  & 0.336\\
 \hline
    7 & 13.857 & 1.384 & 0.909 &  0.274 \\ 
 \hline
   8 & 4.150  & 3.124 &  0.571&0.336  \\
 \hline
   9 & 1.385 & 13.981 & 0.090 & 0.078\\
 \hline
        \end{tabular}
    \end{subtable} 
\end{table}

\begin{table}[!htb]
\setlength\tabcolsep{2pt}
    \caption{Beta distributions' parameter estimates for benign sample type in the PCa dataset under the KN$\cdot$ model.} \label{Table 3}
    \begin{subtable}{0.5\linewidth}
      \centering
        \caption{Patient 1}
 \begin{tabular}{c  c  c c c} 
\hline
 Clusters & $\hat{\alpha}$  & $\hat{\delta}$   & Mean & Std. deviation  \\ 
 \hline
   1 & 13.774 &2.205  &  0.862&0.084  \\ 
 \hline
   2 & 1.491 & 12.454 & 0.107 & 0.080\\
 \hline
   3 & 3.970 & 2.965 & 0.572 & 0.176 \\
 \hline
        \end{tabular}
      \centering
        \caption{Patient 3}
 \begin{tabular}{c  c  c c c} 
  \hline
 Clusters & $\hat{\alpha}$  & $\hat{\delta}$   & Mean & Std. deviation  \\ 
 \hline
   1 & 20.158  & 2.624   & 0.885  &0.065    \\ 
 \hline
   2 & 2.183   & 28.896 & .070  &0.045   \\
 \hline
   3 & 3.618   & 3.023   & 0.545  &0.180   \\
 \hline
        \end{tabular}
    \end{subtable} 
    \hspace*{0.5cm}
     \begin{subtable}{0.5\linewidth}
      \centering
        \caption{Patient 2}
 \begin{tabular}{c  c  c c c} 

   \hline
 Clusters & $\hat{\alpha}$  & $\hat{\delta}$   & Mean & Std. deviation  \\ 
 \hline
   1 & 21.434 &2.871    &0.882   &0.064   \\ 
 \hline
   2 &  2.166  & 18.166  &0.107   &0.067   \\
 \hline
   3 & 4.111   & 2.980   & 0.580  &0.174   \\
 \hline
        \end{tabular}
      \centering
        \caption{Patient 4}
 \begin{tabular}{c  c  c c c} 
 \hline
 Clusters & $\hat{\alpha}$  & $\hat{\delta}$   & Mean & Std. deviation  \\ 
 \hline
   1 & 26.825 & 2.644 & 0.910 &0.052   \\ 
 \hline
   2 & 2.462 & 30.940 &  0.074&0.045 \\
 \hline
   3 & 3.338 &2.237  &0.599  &0.191  \\
 \hline
        \end{tabular}
    \end{subtable} 
\end{table}

\clearpage

\subsection*{Appendix S5 }
\begin{figure*}[h!]
\begin{center}
\includegraphics[width=15cm, height =12cm]{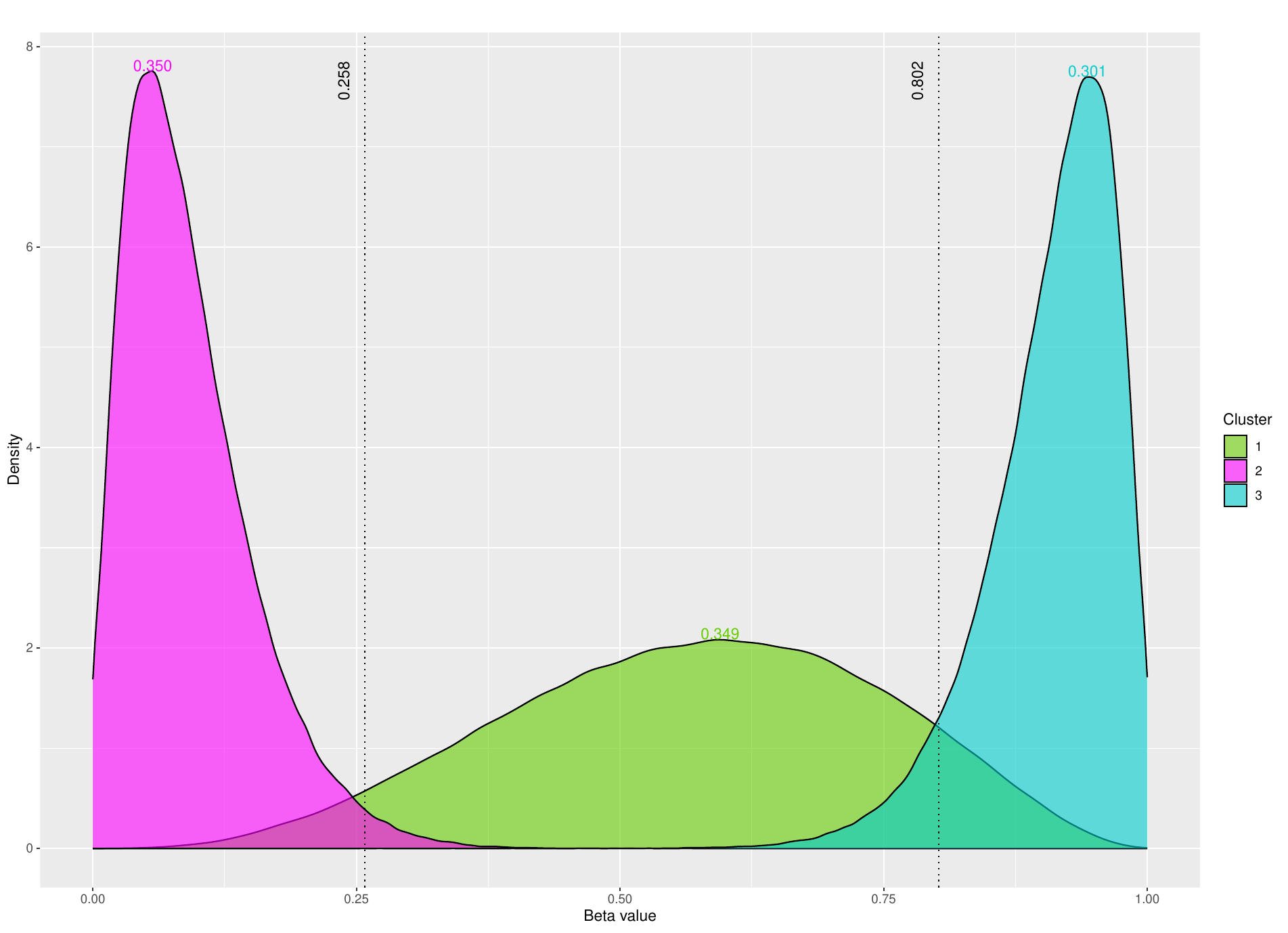}
\caption{Kernel density estimates under the  K$\cdot\cdot$ model fitted to data from  sample type A in the simulated dataset. The thresholds are 0.258 and 0.802. The estimated mixing proportions are displayed.}
\end{center}
\end{figure*}
\clearpage

\subsection*{Appendix S6} 
\begin{figure*}[h!]
\begin{center}
\includegraphics[width=18cm, height =10cm]{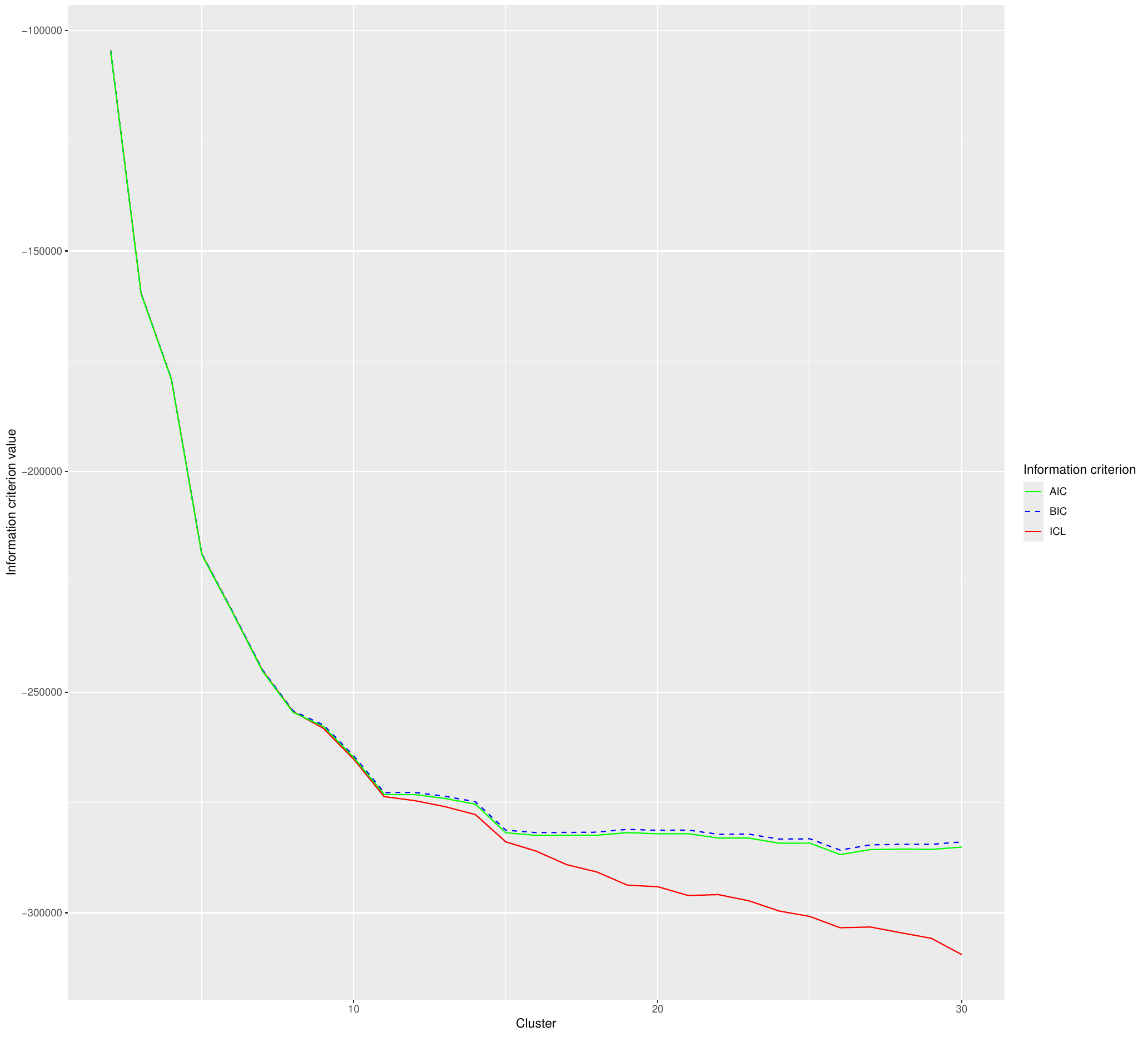}
\caption{The AIC, BIC and ICL information criteria for different numbers of clusters, $K$, for the simulated datasets. 
}
\end{center}
\end{figure*}

\clearpage
\subsection*{Appendix S7 }
\begin{figure*}[h!]
\begin{center}
\includegraphics[width=18cm, height =17cm]{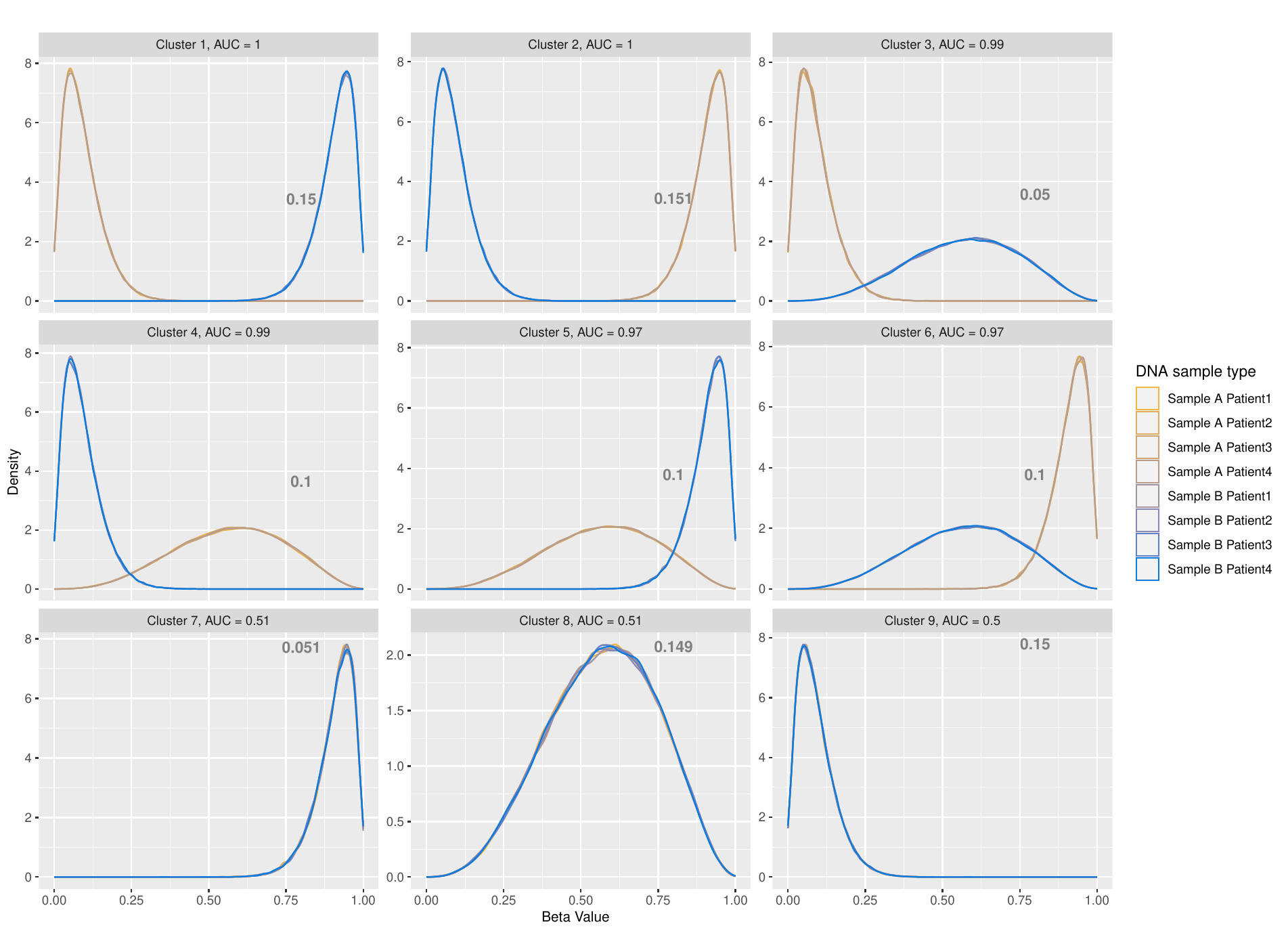}
\caption{Kernel density estimates under the clustering solution of the K$\cdot$R model fitted to DNA samples from sample A and sample B from a simulated dataset. The estimated mixing proportions are  displayed in the relevant panel.}
\end{center}
\end{figure*}

\clearpage
\subsection*{Appendix S8} 
\begin{figure*}[h!]
\begin{center}
\includegraphics[width=18cm, height =14cm]{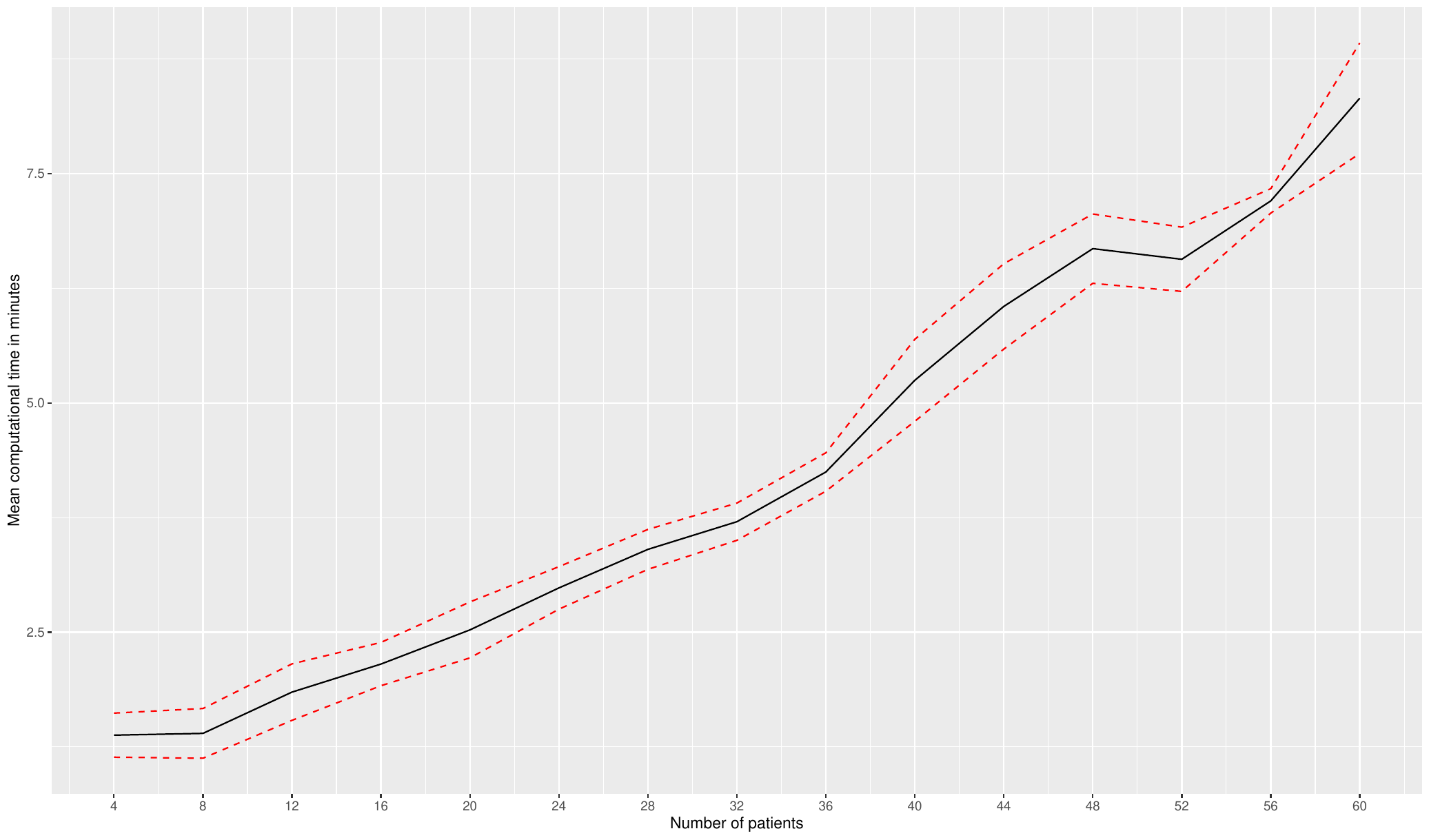}
\caption{Mean computational time for fitting the K$\cdot$R model, with 95\% confidence intervals, as the number of patients $N$ is increased. The computational times for the K$\cdot\cdot$ and KN$\cdot$ models show a similar trend, with elapsed times ranging from 0.33 to 2.5 minutes for the former and 0.47 to 4 minutes for the latter. As the complexity of the algorithm with respect to $N$ is proportional to $N$, as the number of patients increases the computational cost scales linearly.}
\end{center}
\end{figure*}
\clearpage
\subsection*{Appendix S9 }
\begin{figure*}[h!]
\begin{center}
\includegraphics[width=18cm, height =12cm]{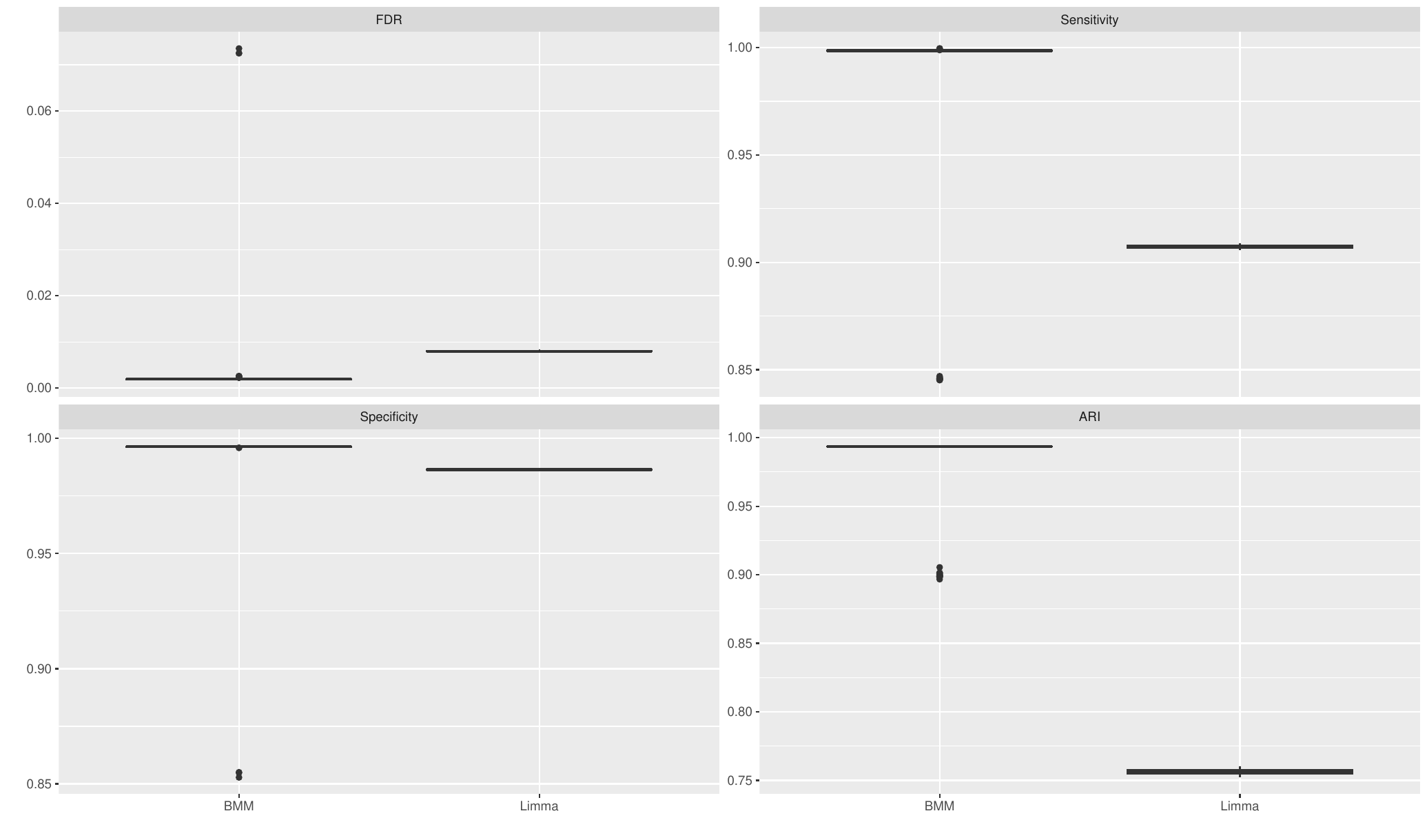}
\caption{Boxplot displaying the FDR, sensitivity, specificity and ARI values from the BMM and Limma methods when applied to the simulated data from a mixture of beta distributions.}
\end{center}
\end{figure*}
\clearpage
\subsection*{Appendix S10}
\begin{figure*}[h!]
\begin{center}
\includegraphics[width=18cm, height =10cm]{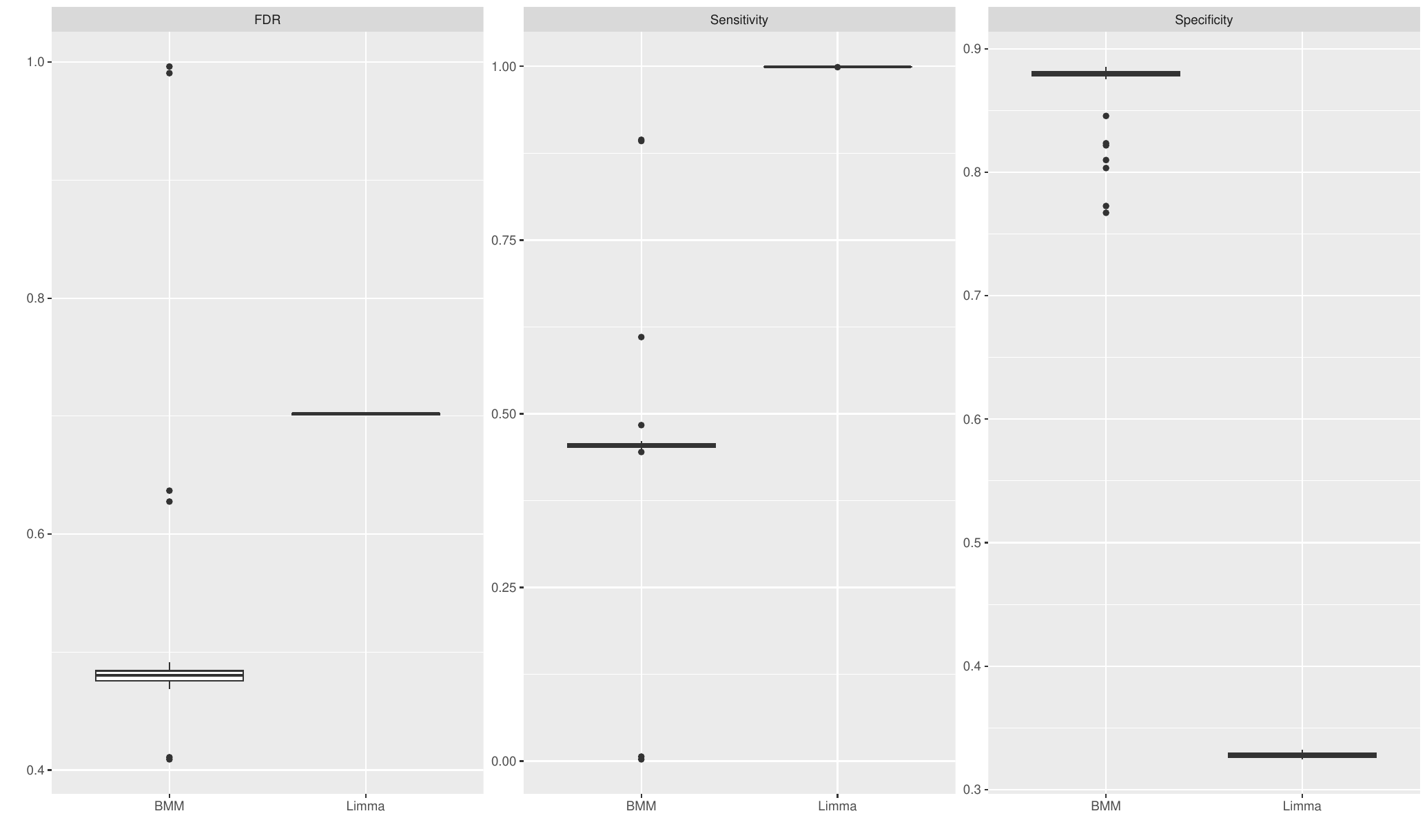}
\caption{Boxplot showing the FDR, sensitivity and specificity values from the BMM and Limma methods applied to the
simulated datasets generated from scaled t-distribution with 8 degrees of freedom to assess the impact of model misspecification.}
\end{center}
\end{figure*}

\clearpage
\subsection*{Appendix S11}
\begin{figure*}[h!]
\begin{center}
\includegraphics[width=0.6\textwidth,height=7cm]{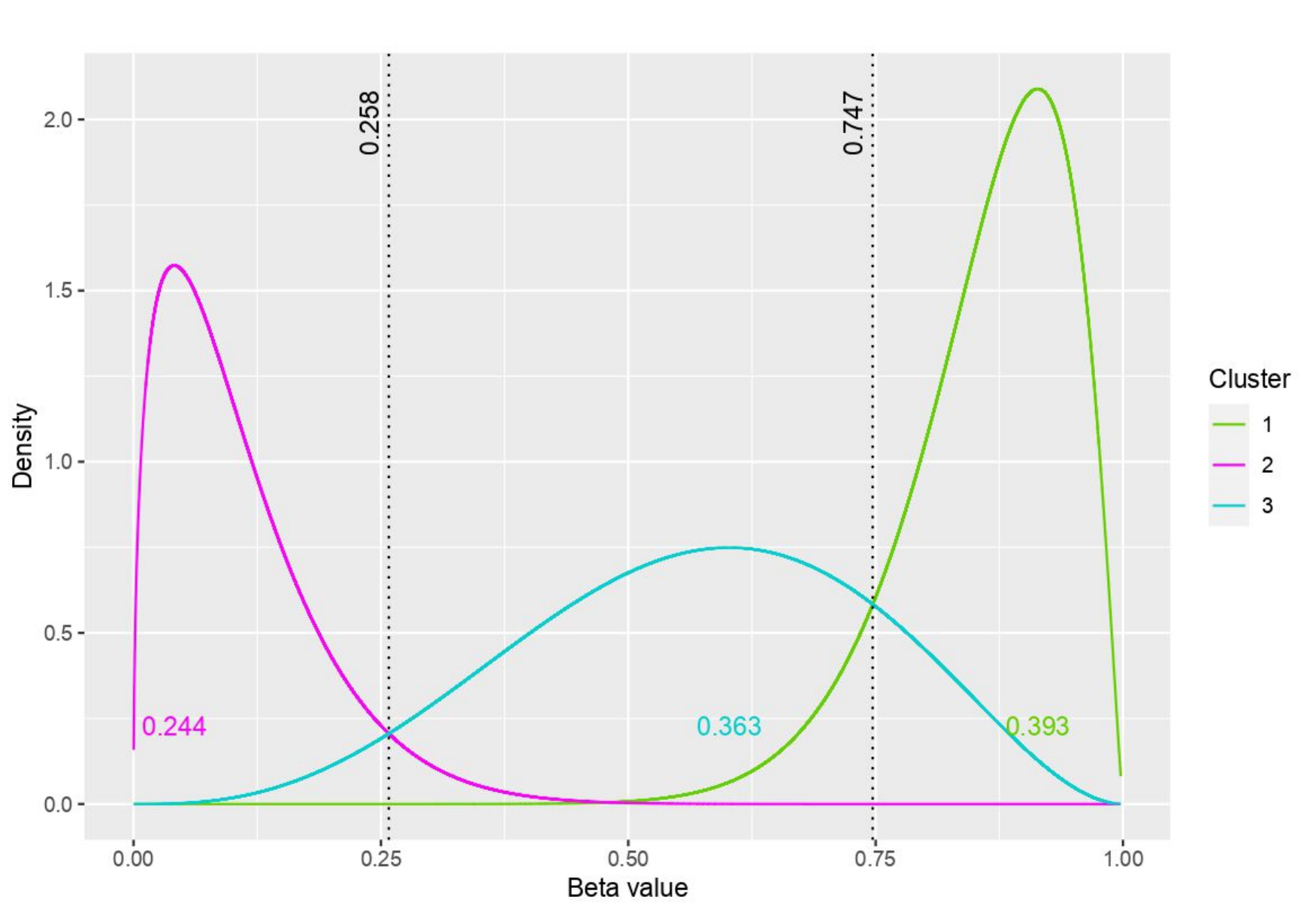}
\caption{Fitted density estimates under the clustering solution of the KN$\cdot$ model fitted to the benign sample collected from patient 1 in the prostate cancer dataset. The threshold points are illustrated in the graph as 0.258 and 0.747.}
\end{center}
\end{figure*}

\begin{figure*}[h!]
\begin{center}
\includegraphics[width=0.6\textwidth,height=7cm]{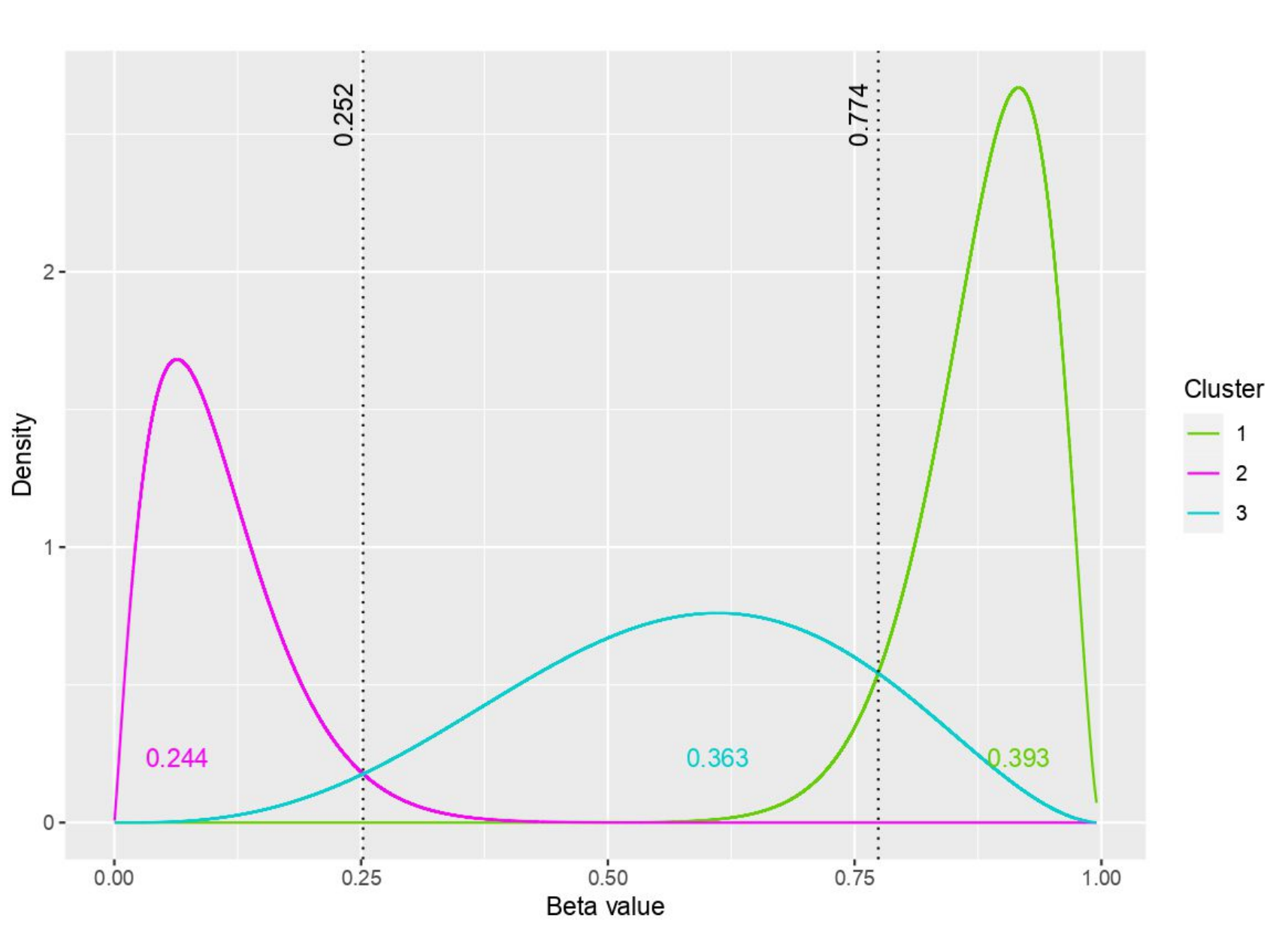}
\caption{Fitted density estimates under the clustering solution of the KN$\cdot$ model fitted to the benign sample collected from patient 2 in the prostate cancer dataset. The threshold points are illustrated in the graph as 0.252 and 0.774.}
\end{center}
\end{figure*}

\begin{figure*}[h!]
\begin{center}
\includegraphics[width=0.6\textwidth,height=7cm]{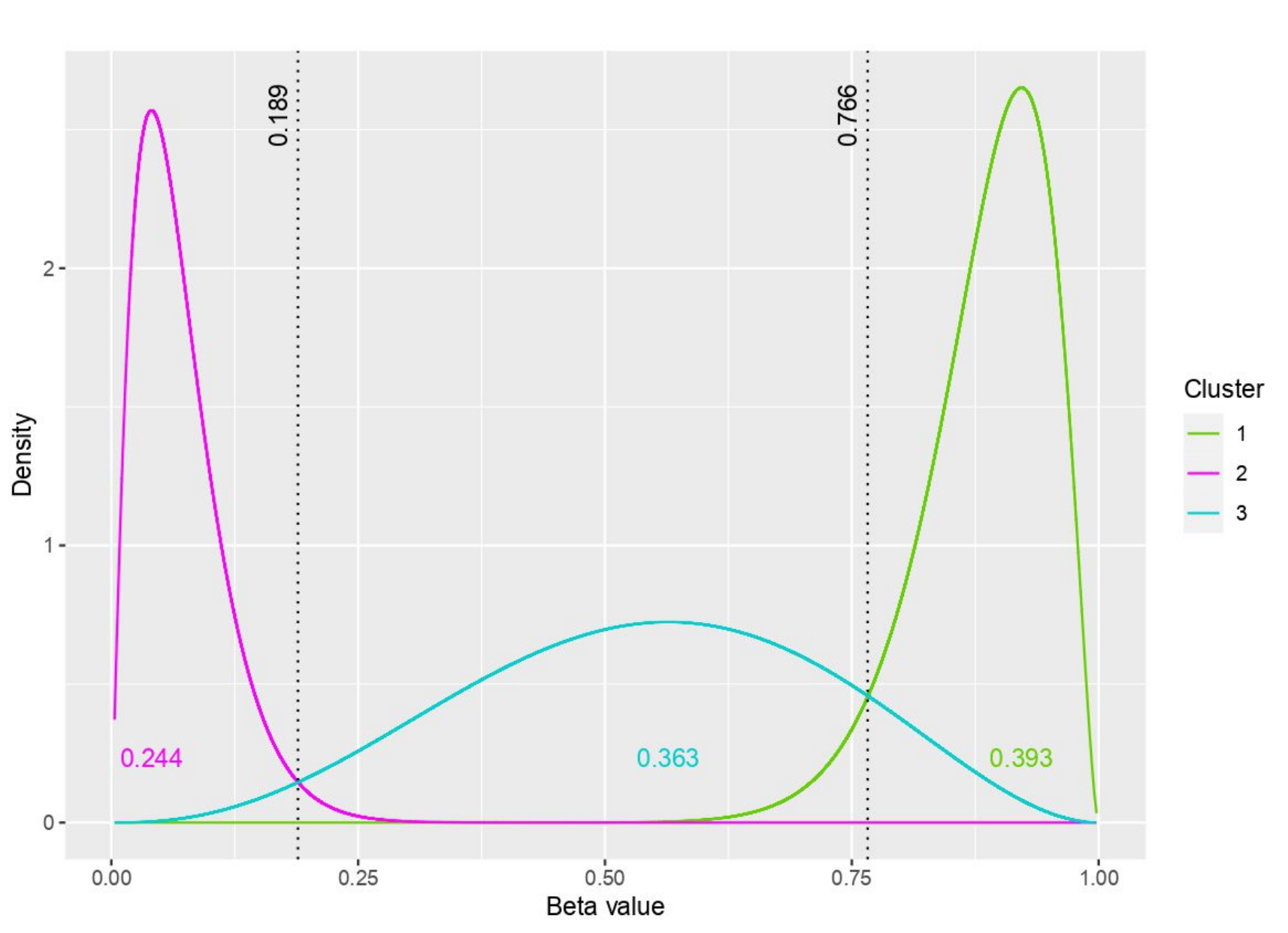}
\caption{Fitted density estimates under the clustering solution of the KN$\cdot$ model fitted to the benign sample collected from patient 3 in the prostate cancer dataset. The threshold points are illustrated in the graph as 0.189 and 0.766.}
\end{center}
\end{figure*}

\begin{figure*}[h!]
\begin{center}
\includegraphics[width=12cm, height =10cm]{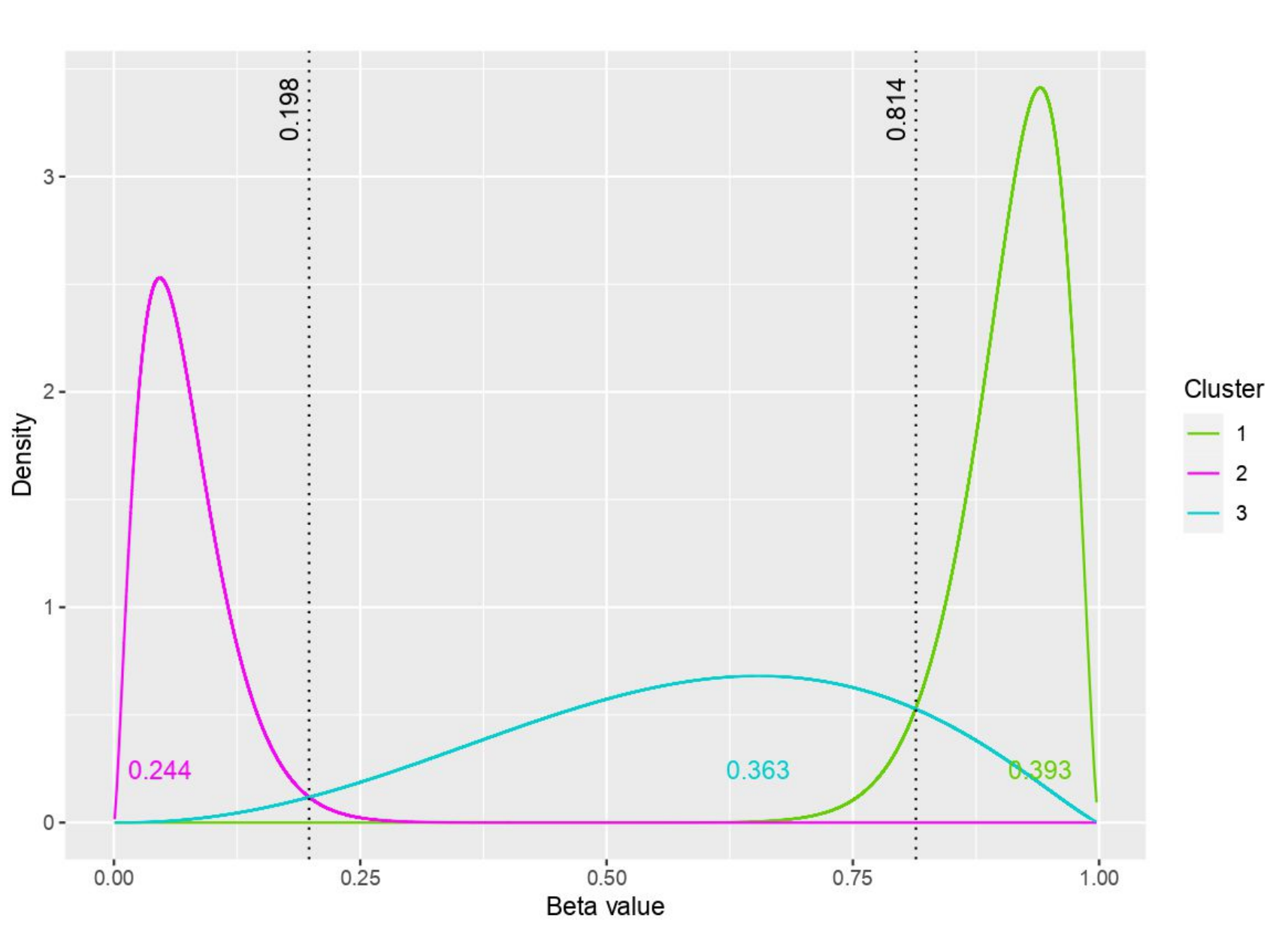}
\caption{Fitted density estimates under the clustering solution of the KN$\cdot$ model fitted to the benign sample collected from patient 4 in the prostate cancer dataset. The threshold points are illustrated in the graph as 0.198 and 0.814.}
\end{center}
\end{figure*}

\begin{figure*}[h!]
\begin{center}
\includegraphics[width=0.6\textwidth,height=8cm]{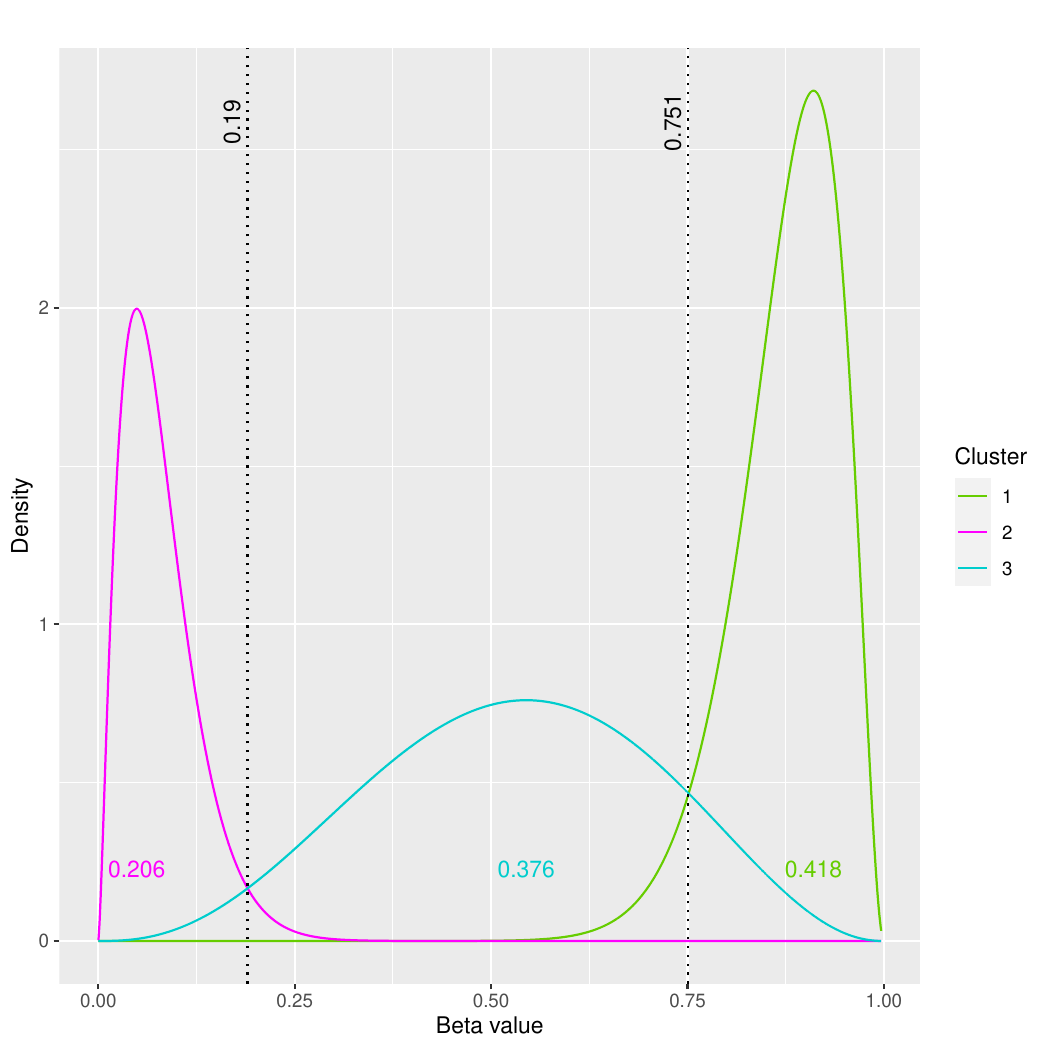}
\caption{Fitted density estimates under the clustering solution of the KN$\cdot$ model fitted to the tumour sample collected from patient 1 in the prostate cancer dataset. The threshold points are illustrated in the graph as 0.19 and 0.751.}
\end{center}
\end{figure*}

\begin{figure*}[h!]
\begin{center}
\includegraphics[width=0.6\textwidth,height=9cm]{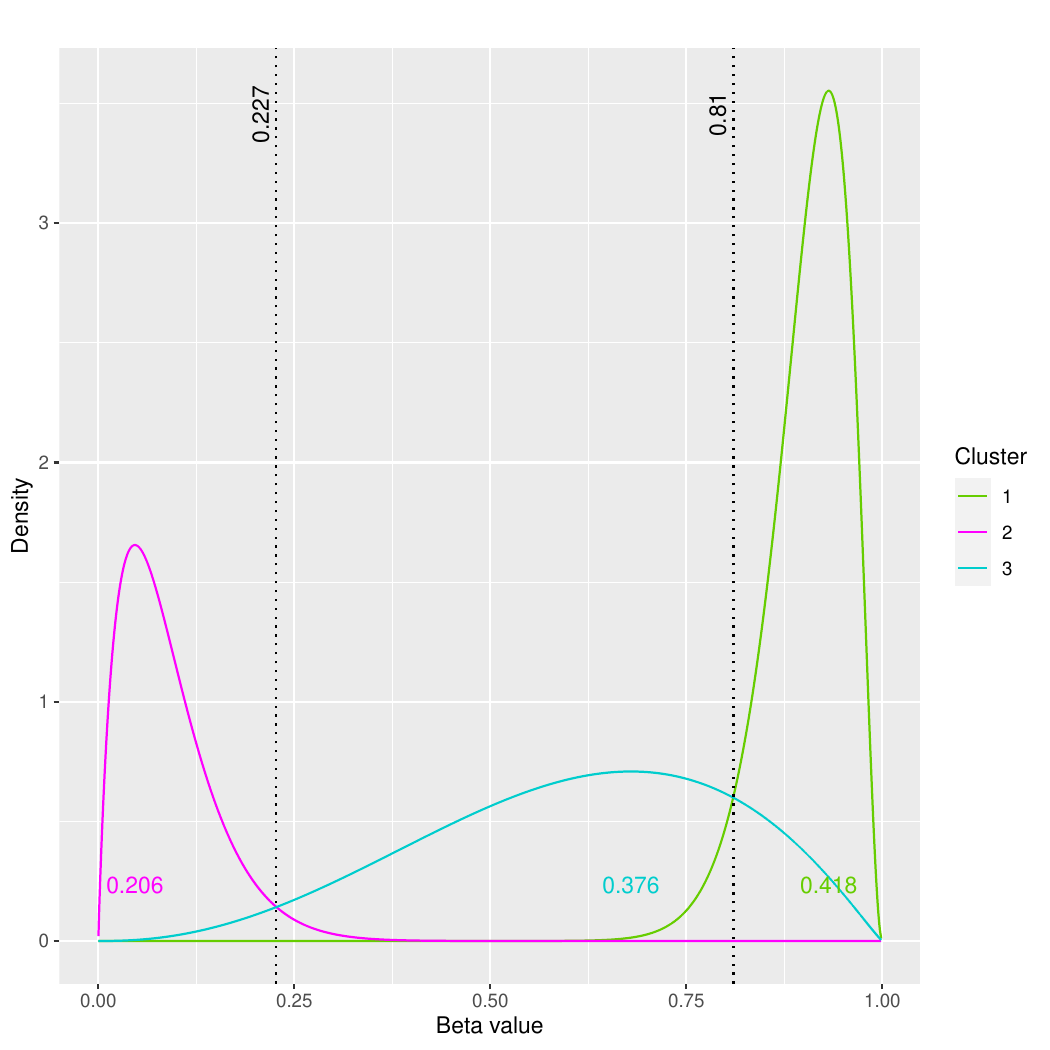}
\caption{Fitted density estimates under the clustering solution of the KN$\cdot$ model fitted to the tumour sample collected from patient 2 in the prostate cancer dataset. The threshold points are illustrated in the graph as 0.227 and 0.81.}
\end{center}
\end{figure*}

\begin{figure*}[h!]
\begin{center}
\includegraphics[width=0.6\textwidth,height=8cm]{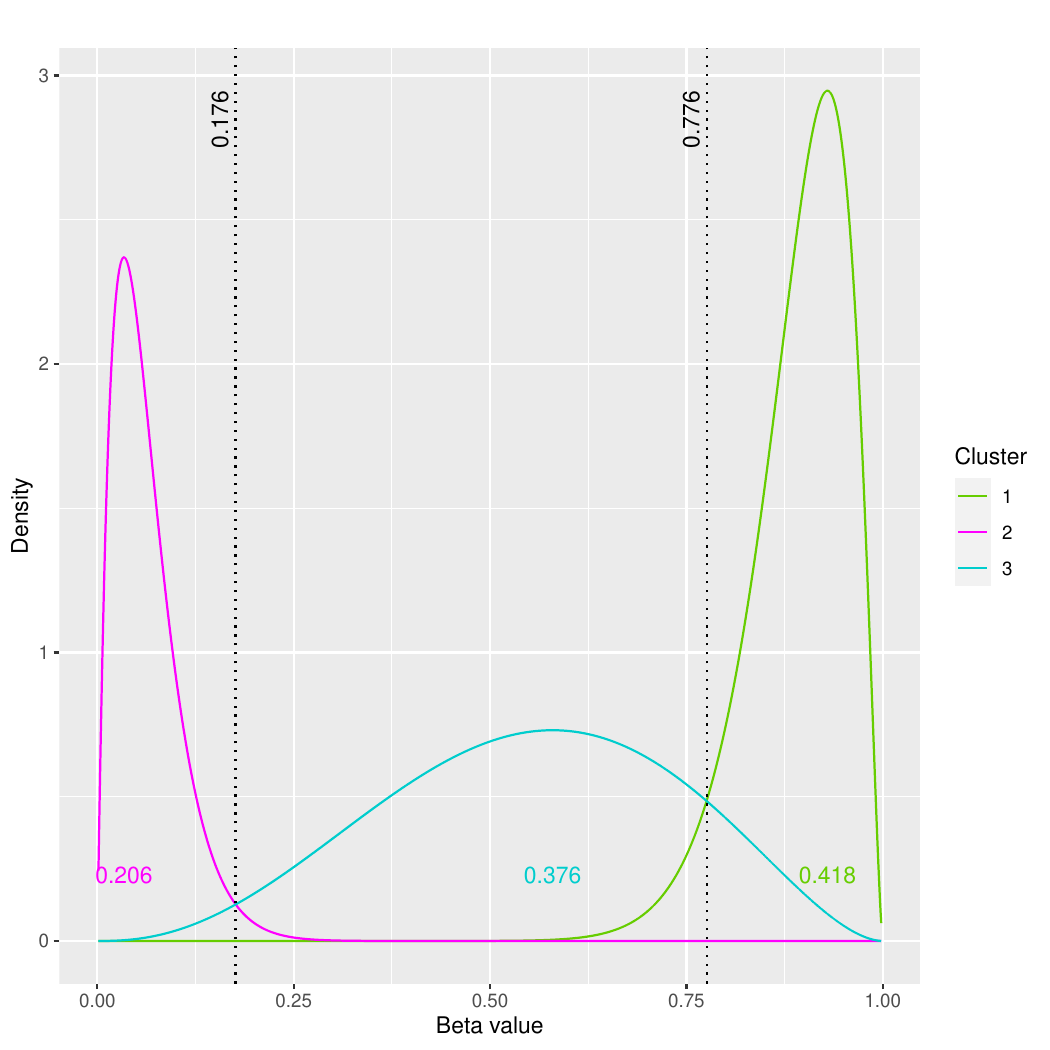}
\caption{Fitted density estimates under the clustering solution of the KN$\cdot$ model fitted to the tumour sample collected from patient 3 in the prostate cancer dataset. The threshold points are illustrated in the graph as 0.176 and 0.776.}
\end{center}
\end{figure*}

\begin{figure*}[h!]
\begin{center}
\includegraphics[width=12cm, height =9cm]{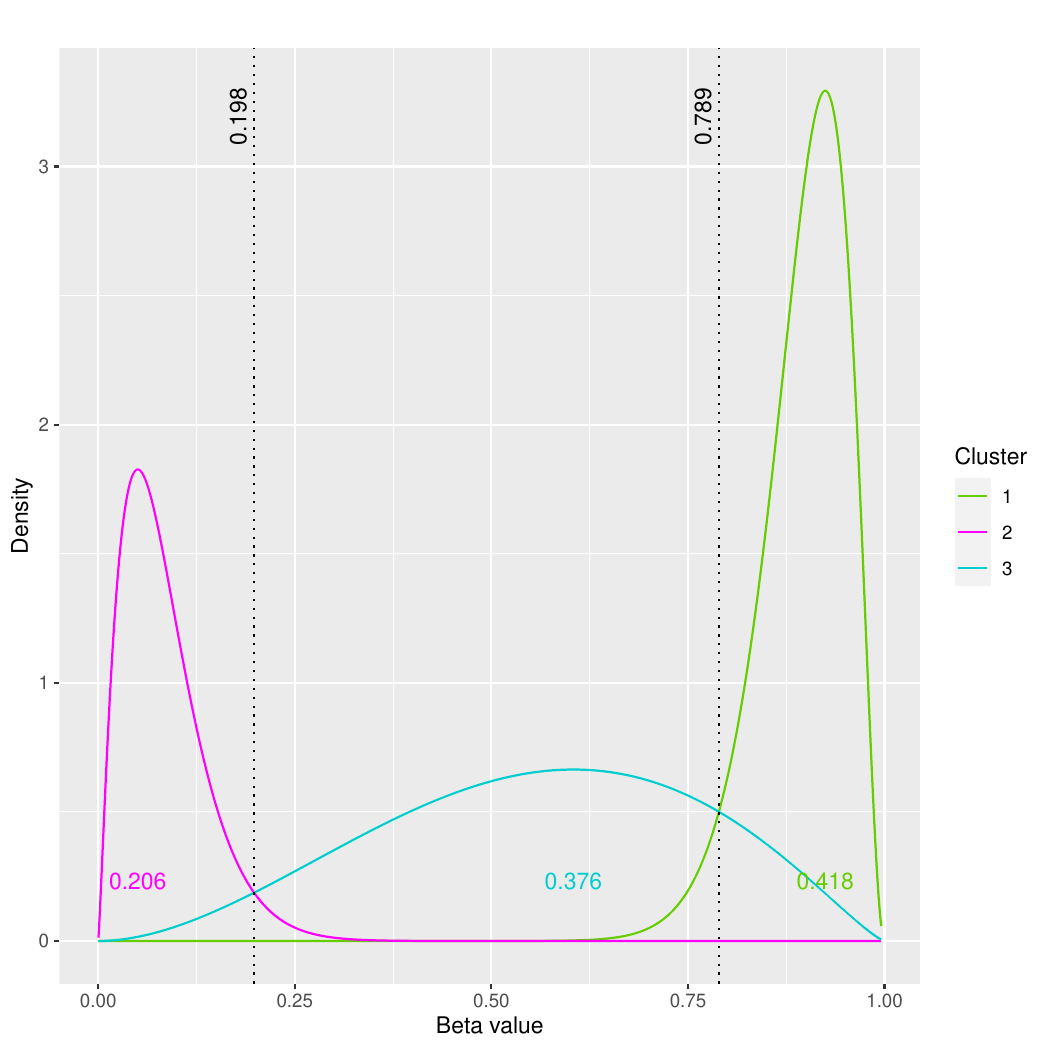}
\caption{Fitted density estimates under the clustering solution of the KN$\cdot$ model fitted to the tumour sample collected from patient 4 in the prostate cancer dataset. The threshold points are illustrated in the graph as 0.198 and 0.789.}
\end{center}
\end{figure*}

\clearpage
\subsection*{Appendix S12}
\begin{figure*}[h!]
\begin{center}
\includegraphics[width=15cm, height =12cm]{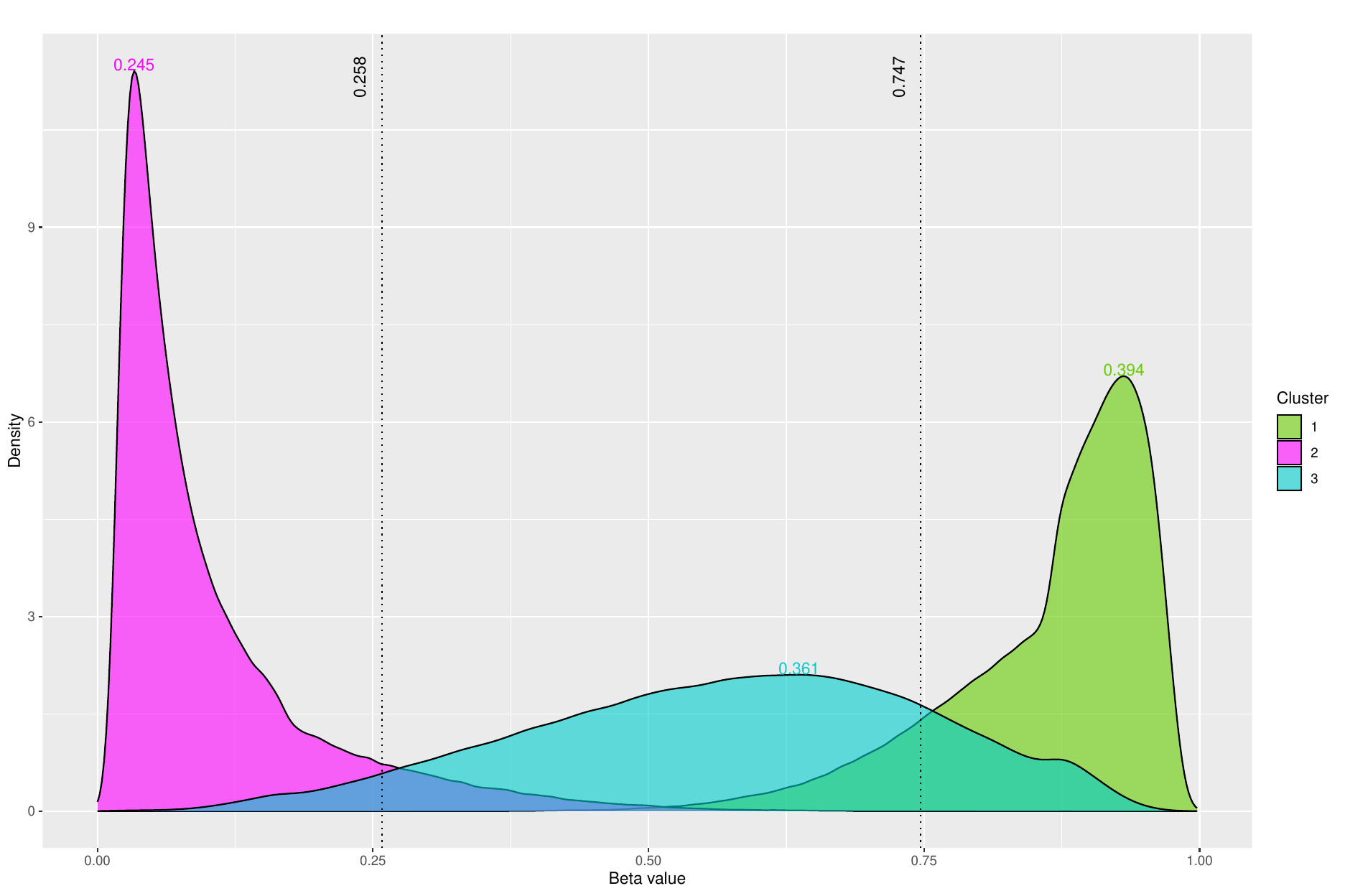}
\caption{Kernel density estimates under the clustering solution of the KN$\cdot$ model fitted to DNA methylation data from the benign sample collected from patient 1 in the prostate cancer dataset. The thresholds are illustrated along with the estimated mixing proportions.}
\end{center}
\end{figure*}
\clearpage

\subsection*{Appendix S13} 
\begin{figure*}[h!]
\begin{center}
\includegraphics[width=18cm, height =12cm]{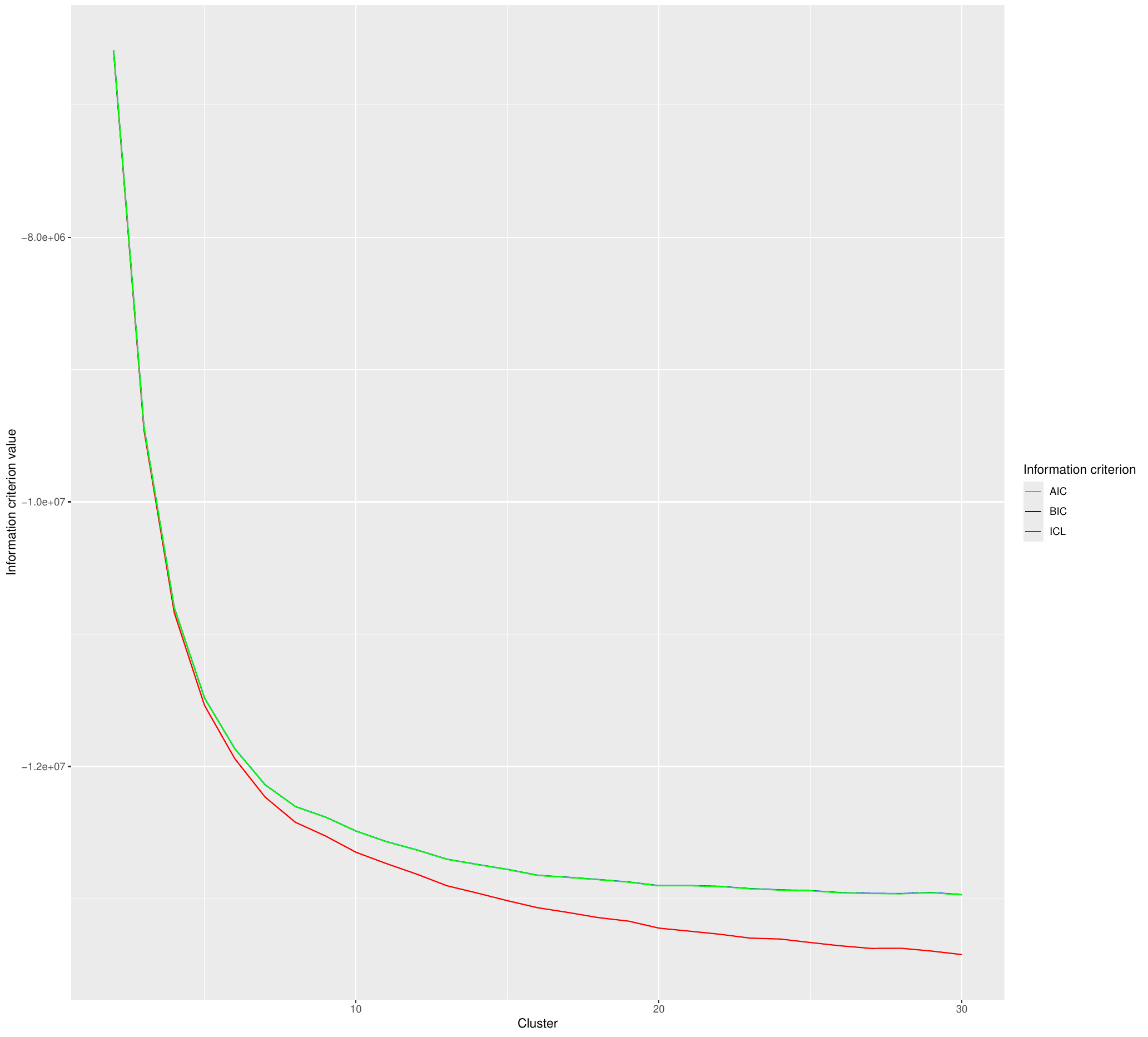}
\caption{The AIC, BIC and ICL information criteria for different numbers of clusters, $K$, for the PCa dataset. 
}
\end{center}
\end{figure*}

\clearpage
\subsection*{Appendix S14 }
\begin{figure*}[h!]
\begin{center}
\includegraphics[width=18cm, height =15cm]{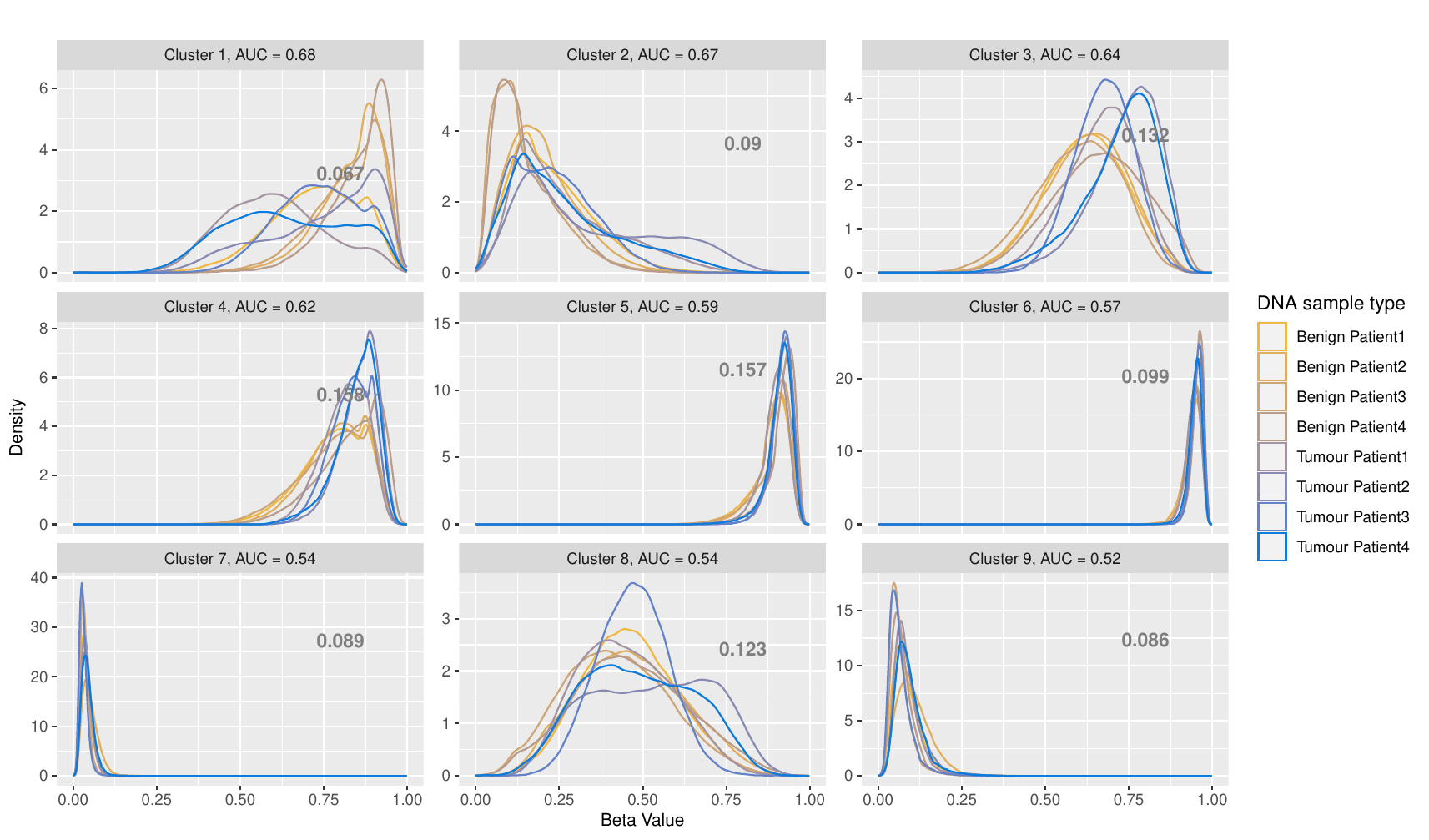}
\caption{Kernel density estimates under the clustering solution of the K$\cdot$R model fitted to the DNA methylation data from benign and tumour prostate cancer samples. The estimated mixing proportions are displayed in the relevant panel.}
\end{center}
\end{figure*}

\clearpage
\subsection*{Appendix S15}

\begin{figure*}[h!]
\begin{center}
\includegraphics[width=0.65\textwidth,height=9.5cm]{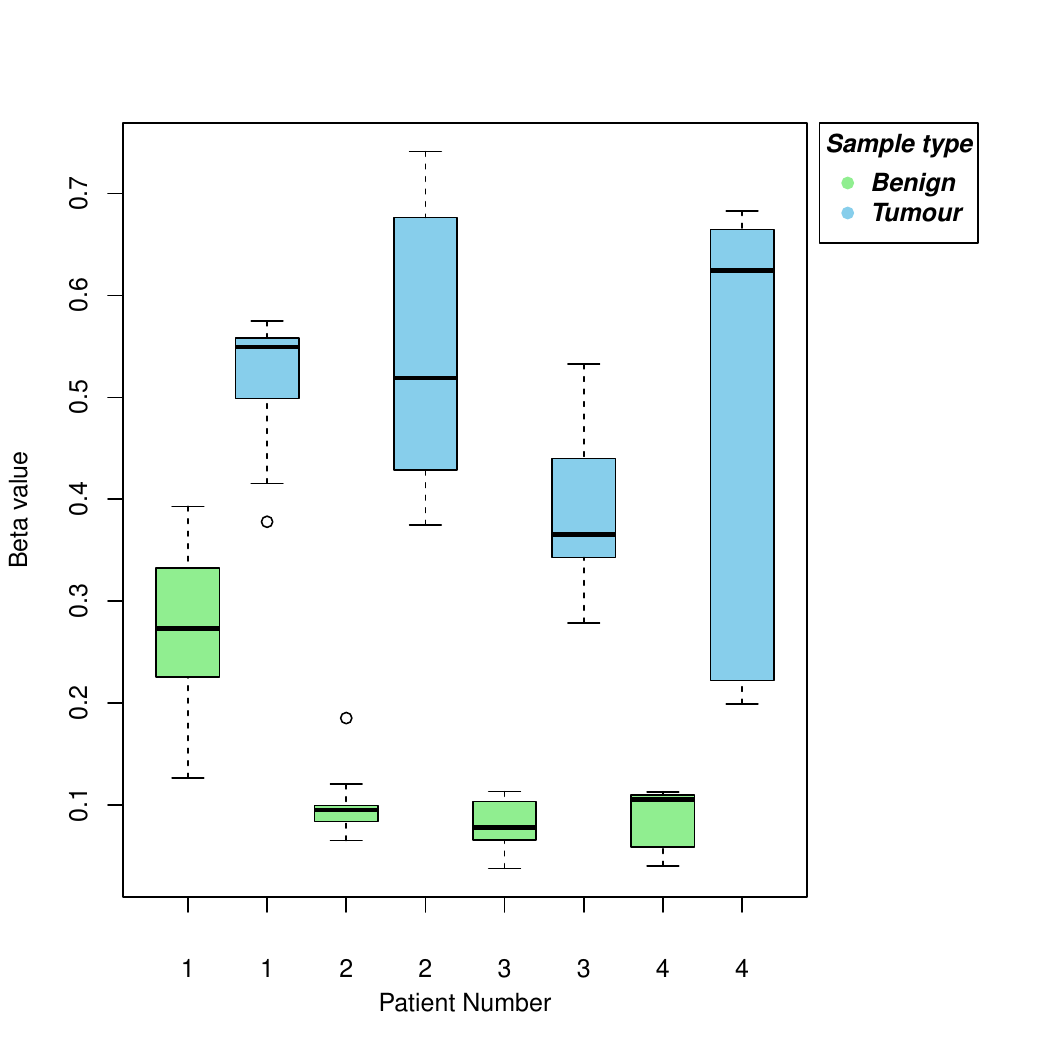}
\caption{Methlyation levels of the differentially methylated CpG sites related to the RARB genes in the benign and tumour sample types.}
\end{center}
\end{figure*}

\clearpage
\subsection*{Appendix S16 }

\begin{figure*}[h!]
\begin{center}
\includegraphics[width=0.7\textwidth,height=9.5cm]{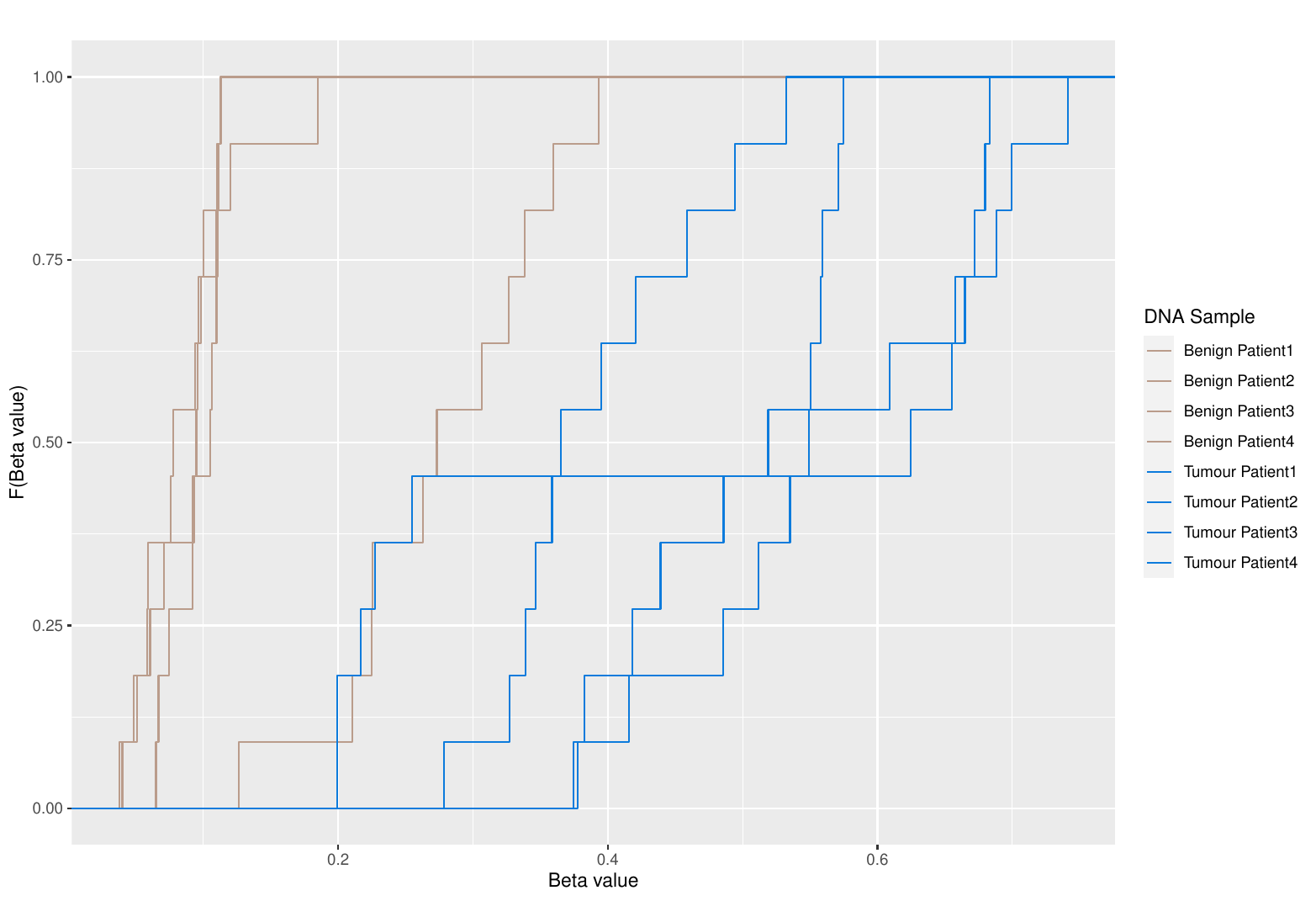}
\caption{ECDFs for the DMCs related to the RARB genes for all patients and sample types.}
\end{center}
\end{figure*}

\clearpage
\subsection*{Appendix S17}
\begin{figure*}[h!]
\begin{center}
\includegraphics[width=0.65\textwidth,height=8cm]{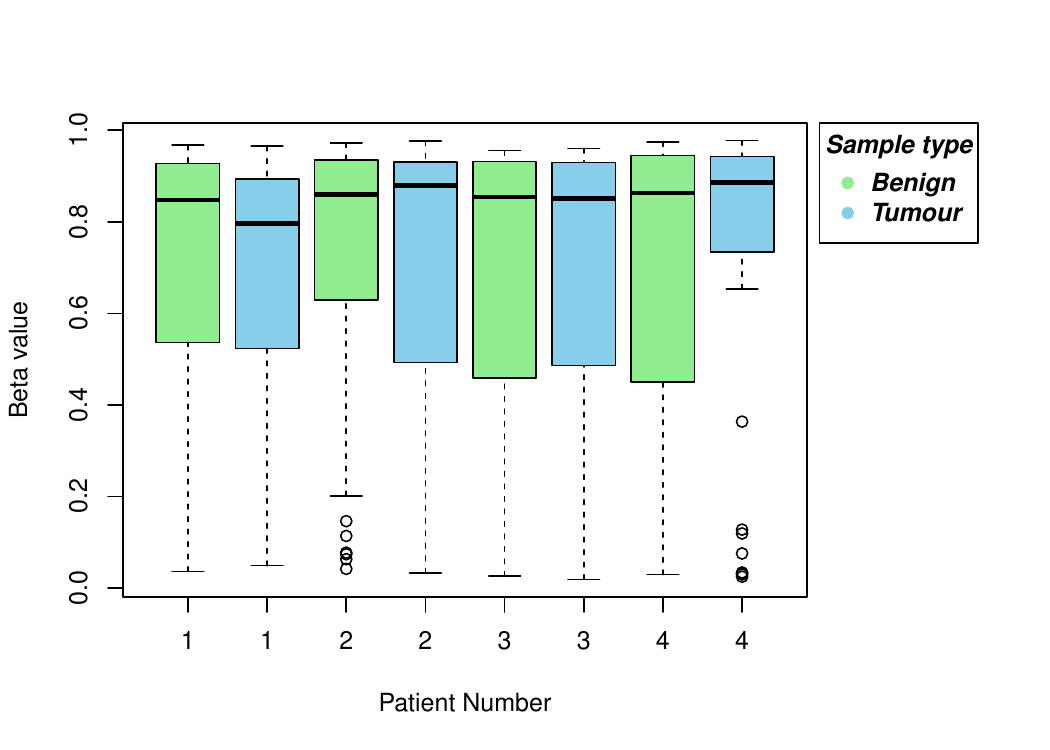}
\caption{Methylation levels of the differentially methylated CpG sites in clusters 3-9  related to the AKT1 gene for all patients and sample types.}
\end{center}
\end{figure*}

\begin{figure*}[h!]
\begin{center}
\includegraphics[width=0.8\textwidth, height =9cm]{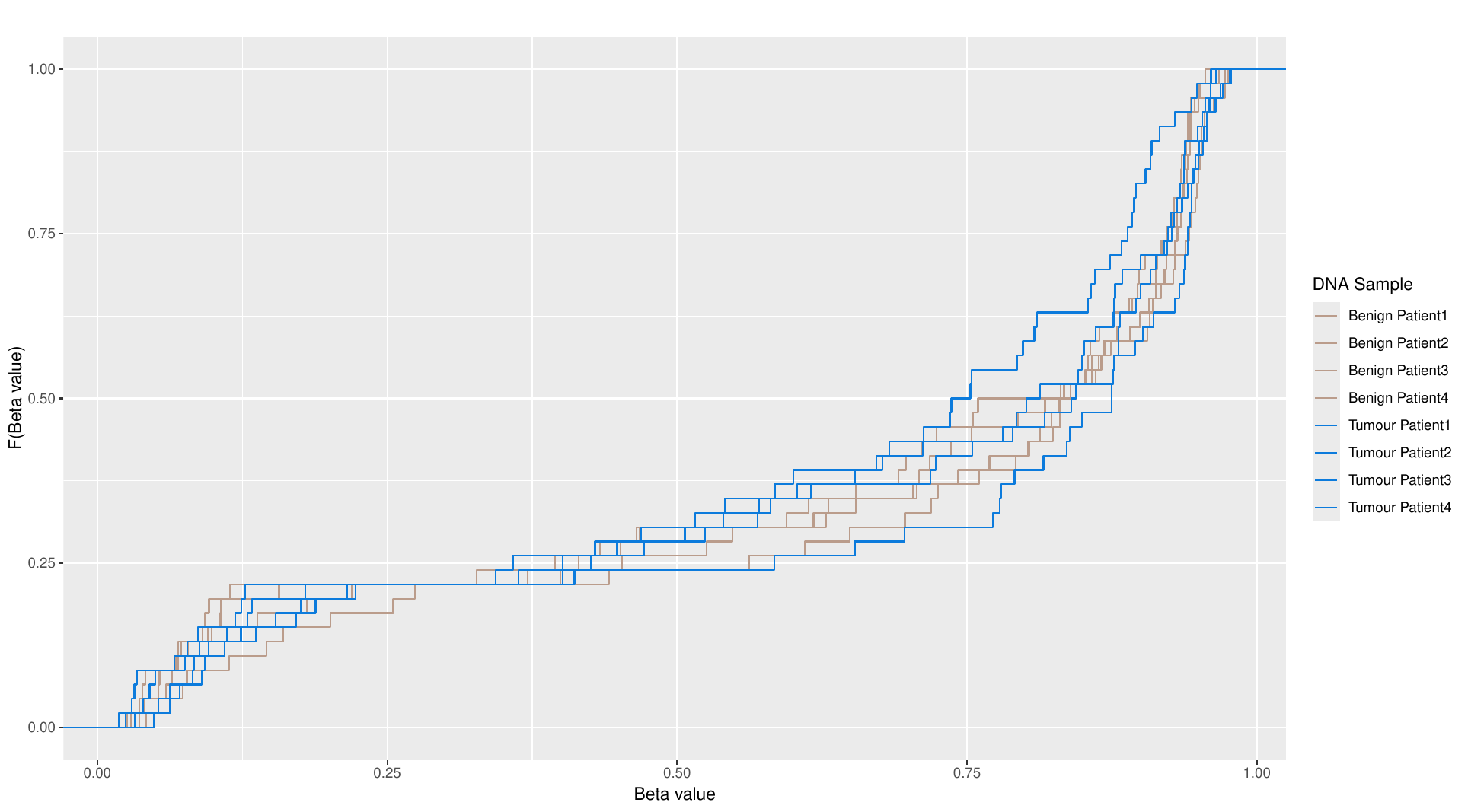}
\caption{ECDFs for the CpG sites in clusters 3-9  related to the AKT1 gene for all patients and sample types.}
\end{center}
\end{figure*}

\clearpage
\subsection*{Appendix S18}
\begin{figure*}[h!]
\begin{center}
\includegraphics[width=0.8\textwidth, height =9cm]{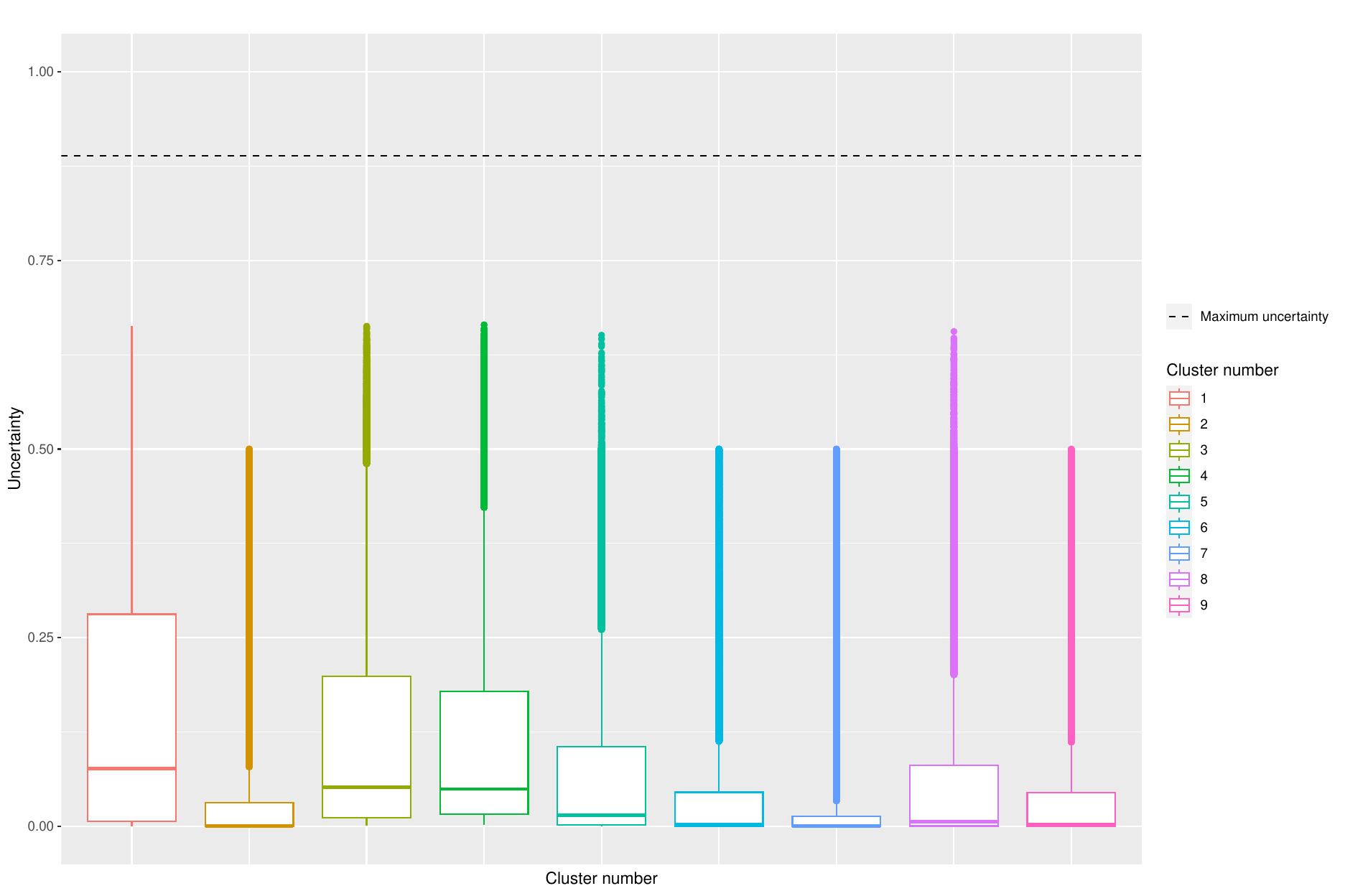}
\caption{Clustering uncertainties for CpG sites in the PCa data.}
\end{center}
\end{figure*}

 \clearpage

\subsection*{Appendix S19}
\subsubsection*{Esophageal squamous cell carcinoma data}

Esophageal squamous cell carcinoma (ESCC) is a subtype of esophageal cancer characterized by aberrant DNA methylation. A study was conducted to investigate abnormal genes in ESCC, and DNA samples were collected from 15 patients' benign and tumour tissues \citep{ESCC}. Paired samples from 4 randomly selected patients were considered. Approximately $1.34$ \% of the methylation values were $0$; given this low prevalence, here $0$ values were substituted with the minimum observed \textit{beta} value.  The ESCC dataset  contained observed \textit{beta} values for $C = 481,315$  CpG sites from each of $R =$ 2 DNA sample types collected from each of $N =$ 4 patients. These data were accessed from \href{https://www.ncbi.nlm.nih.gov/geo/query/acc.cgi?acc=GSE121931}{GEO repository(GSE121931)} on 23\textsuperscript{rd} of February, 2022 for research purposes. The authors had no access to information that could identify individual participants.

\subsubsection*{Estimating methylation state thresholds}
\begin{figure}[h!]
\begin{center}
\includegraphics[width=1\textwidth,height=9cm]{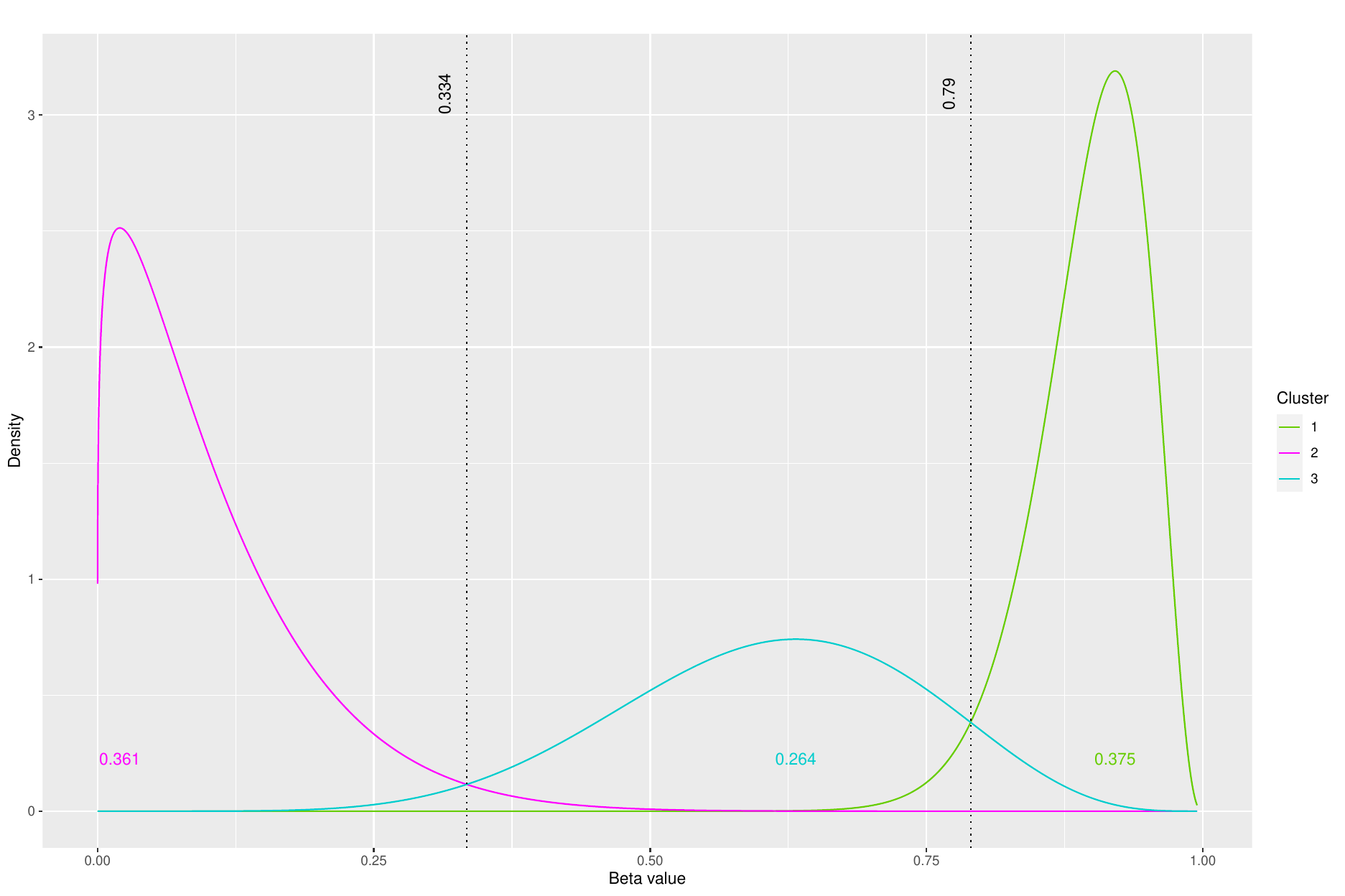}
\caption{Fitted density estimates under the clustering solution of the KN$\cdot$ model for DNA methylation data from the benign sample collected from patient 1 in the ESCC dataset. The methylation state thresholds are illustrated by the black dotted lines along with the estimated mixing proportions.}
\label{ESCC:Figure1}
\end{center}
\end{figure}

The methylation states of each CpG site and the threshold points between these states are to be inferred. The K$\cdot\cdot$ and KN$\cdot$ models are used to achieve this objective by clustering the CpG sites from the benign sample into 3 methylation states, allowing objective inference of the thresholds. The KN$\cdot$ model was selected as the optimal model by BIC and the fitted density estimates of the clustering solution for patient 1 are displayed in Figure \ref{ESCC:Figure1}. As the KN$\cdot$ model estimates different parameters for each patient, different pairs of thresholds are calculated for each patient.
The methylation state thresholds for patient 1 are inferred to be 0.334 and 0.79 under the KN$\cdot$ model. A summary of the parameter estimates under the KN$\cdot$ model is presented in Table \ref{ESCC:Table 4}.

\begin{table}[!htb]
\setlength\tabcolsep{2pt}

    \caption{Beta distributions' parameter estimates for benign samples in the ESCC dataset under the KN$\cdot$ model.} \label{ESCC:Table 4}
    \begin{subtable}{0.5\linewidth}
      \centering
        \caption{Patient 1}
 \begin{tabular}{c  c  c c c} 
 \hline
 Clusters & $\hat{\alpha}$  & $\hat{\delta}$   & Mean & Std. deviation  \\ 
 \hline
   1 & 31.640 & 3.642  &  0.897 &0.051  \\ 
 \hline
   2 & 1.193 & 10.385 & 0.103 & 0.086\\
 \hline
   3 & 7.369 & 4.713 & 0.610 & 0.135 \\
 \hline
        \end{tabular}
      \centering
        \caption{Patient 3}
 \begin{tabular}{c  c  c c c} 
 \hline
 Clusters & $\hat{\alpha}$  & $\hat{\delta}$   & Mean & Std. deviation  \\ 
 \hline
   1 & 28.449  & 3.880   & 0.880  &0.056    \\ 
 \hline
   2 & 1.278   & 10.7 & 0.107  &0.086   \\
 \hline
   3 & 7.509   & 5.005   & 0.600  &0.133   \\
 \hline
        \end{tabular}
    \end{subtable} 
    \hspace*{0.5cm}
     \begin{subtable}{0.5\linewidth}
      \centering
        \caption{Patient 2}
 \begin{tabular}{c  c  c c c} 

  \hline
 Clusters & $\hat{\alpha}$  & $\hat{\delta}$   & Mean & Std. deviation  \\ 
 \hline
   1 & 31.532 &3.632    &0.897   &0.051   \\ 
 \hline
   2 &  1.162  & 10.587  &0.099   &0.084   \\
 \hline
   3 & 6.600  & 4.274   & 0.607  &0.142  \\
 \hline
        \end{tabular}
      \centering
        \caption{Patient 4}
 \begin{tabular}{c  c  c c c} 
 \hline
 Clusters & $\hat{\alpha}$  & $\hat{\delta}$   & Mean & Std. deviation  \\ 
 \hline
   1 & 29.519 & 3.986 & 0.881 &0.055   \\ 
 \hline
   2 & 1.281 & 9.552 &  0.118& 0.094\\
 \hline
   3 & 7.509 & 4.948  &0.593 & 0.135 \\
 \hline
        \end{tabular}
    \end{subtable} 
\end{table}

\subsubsection *{Identifying DMCs in the ESCC data}

\begin{figure*}[h!]
\begin{center}
\includegraphics[width=18cm, height =15cm]{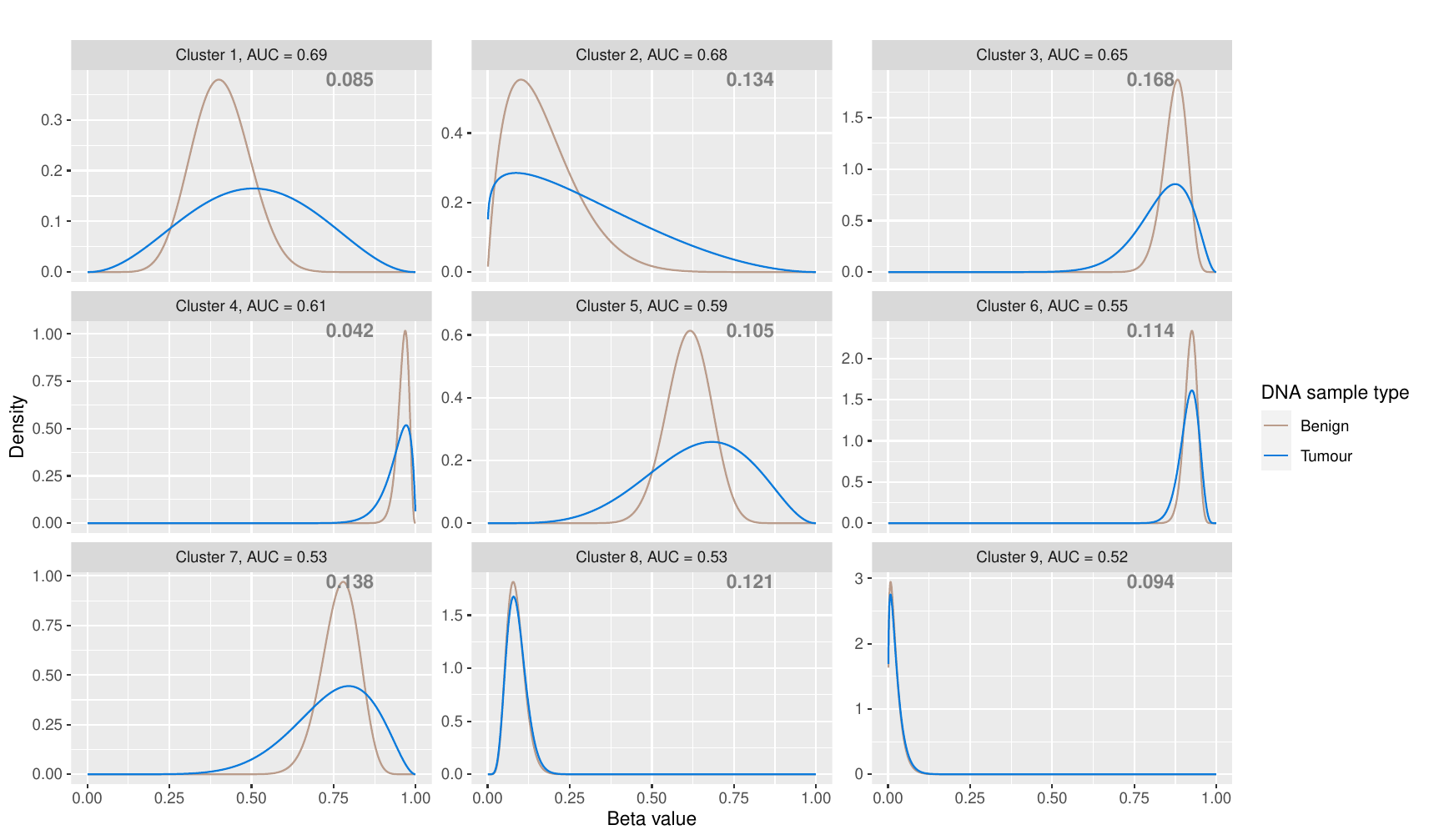}
\caption{Fitted density estimates under the clustering solution of the K$\cdot$R model for DNA methylation data from benign and tumour ESCC samples. The estimated mixing proportions are displayed in the relevant panel.}
\label{ESCC:Figure2}
\end{center}
\end{figure*}

The CpG sites that are differentially methylated between the benign and tumour samples are identified by fitting the K$\cdot$R model with biologically motivated $K = 9$ clusters. The fitted densities are shown in Figure \ref{ESCC:Figure2}. 
A summary of the parameter estimates under the K$\cdot$R model is presented in Table \ref{ESCC:Table 5}.
\begin{table}[!htb]
    \caption{Beta distributions' parameter estimates for the ESCC dataset under the K$\cdot$R model.}. \label{ESCC:Table 5}
    \setlength\tabcolsep{2.4pt}
    \begin{subtable}{.5\linewidth}
      \centering
        \caption{Benign samples}
 \begin{tabular}{c  c  c c c} 
 \hline
 Clusters & $\hat{\alpha}$  & $\hat{\delta}$   & Mean & Std. deviation  \\ 
 \hline
   1  &12.508     & 18.222    &0.407  & 0.087 \\ 
 \hline
   2 &1.968     &   9.562   &  0.171 &0.106  \\
 \hline
   3 & 72.289   & 10.548  & 0.873 & 0.036  \\
 \hline
    4  & 114.017   & 4.712    &0.960   &  0.018 \\ 
 \hline
   5&31.205    & 19.781   &0.612   &0.068  \\
 \hline
   6  & 170.783  & 14.678    & 0.921  &  0.020\\
 \hline
    7  & 41.830   & 12.567  & 0.769  & 0.057  
      \\ 
 \hline
   8 & 8.884    & 94.511  & 0.086  & 0.027  \\
 \hline
   9 & 1.496    & 63.762   & 0.023  &  0.018 \\
 \hline
        \end{tabular}
        
    \end{subtable}
    \hspace*{0.5cm}
    \begin{subtable}{.5\linewidth}
      \centering
        \caption{Tumour samples}
 \begin{tabular}{c  c  c c c} 
 \hline
  Clusters & $\hat{\alpha}$  & $\hat{\delta}$   & Mean & Std. deviation  \\ 
 \hline
   1  & 3.230 & 3.181 &0.504  & 0.184 \\ 
 \hline
   2 & 1.179 & 2.897 & 0.289 & 0.201 \\
 \hline
   3 & 16.145 &3.174  &  0.836& 0.082 \\
 \hline
    4 & 30.385 & 1.876 & 0.942 & 0.041 \\ 
 \hline
   5 & 5.679 & 3.176 & 0.641 &  0.153  \\
 \hline
   6 & 82.000 & 7.538 & 0.916 & 0.029 \\
 \hline
    7& 8.581 & 2.944 & 0.745 & 0.123
     \\ 
 \hline
   8 &  7.922& 81.721 &0.088  &  0.030\\
 \hline
   9 & 1.421 & 56.256 &0.025  & 0.020 \\
 \hline
        \end{tabular}
    \end{subtable} 
\end{table}

\begin{figure*}[h!]
\begin{center}
\includegraphics[width=1\textwidth, height =12cm]{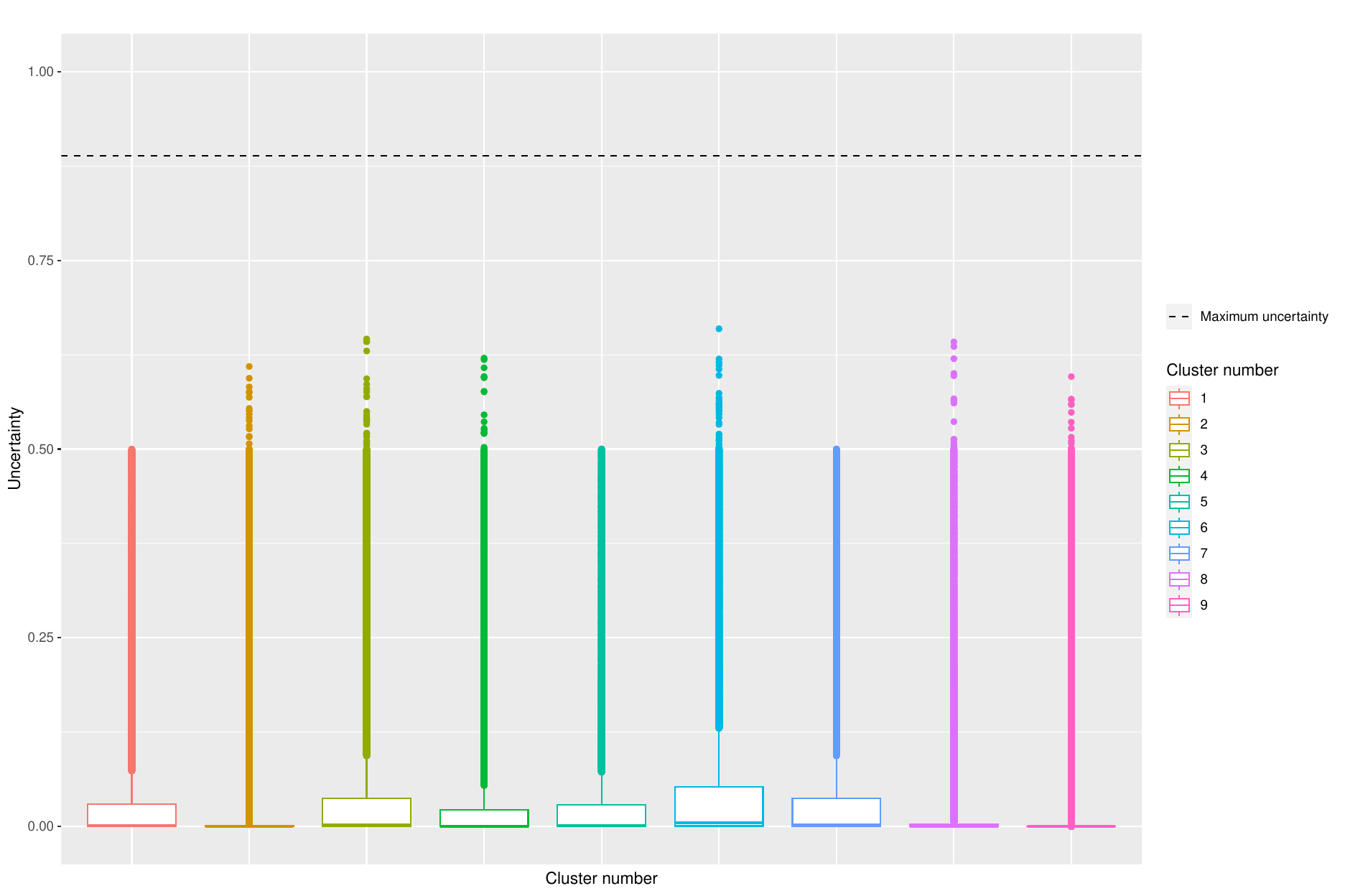}
\caption{Clustering uncertainties for CpG sites in each clustering group under the clustering solution of the K·R model for the ESCC dataset.}
\label{ESCC:Figure3}
\end{center}
\end{figure*}

The maximum possible uncertainty when clustering the CpG sites into $K$ clusters is $1-1/K = 8/9$. Figure \ref{ESCC:Figure3} illustrates the clustering uncertainties for all CpG sites and demonstrates that there is low uncertainty in the CpG site's cluster memberships under the K$\cdot$R model. We subsequently utilized the AUC and WD metrics to quantify the dissimilarity among the cumulative distributions within each cluster. The clusters are illustrated in the descending order of their degree of differential methylation, based on decreasing AUC and, in the case of ties, WD values. These metrics and the parameter estimates suggest clusters 1 and 2 to be the most differentially methylated clusters thus identifying 104,930 CpG sites as being the most differentially methylated CpG sites. Detailed AUC and WD metric values for each individual cluster can be found in Table \ref{ESCC:Table 6}.

 \begin{table}
 \centering
 \caption{The AUC and WD metrics calculated across each cluster for the ESCC dataset.}
\label{ESCC:Table 6}
 \begin{tabular}{|l |c | c |c|c|c|c|c|c|c|} 
\cline{2-10}
 \multicolumn{1}{l|}{} & \multicolumn{9}{|c|}{Cluster} \\
 \cline{2-10}
 \multicolumn{1}{c|}{ } & 1 & 2 & 3 & 4 & 5 & 6 & 7 & 8 & 9 \\ 
 \hline
  AUC& 0.691 &0.679& 0.650& 0.611& 0.588& 0.547& 0.535& 0.529& 0.524 \\ 

 \hline
 WD   & 0.114 &0.119& 0.043 &0.020& 0.076 &0.008 &0.054& 0.003& 0.002  \\
 \hline

\end{tabular}
\end{table} 

\begin{figure*}
\begin{center}
\includegraphics[width=14cm, height= 9cm]{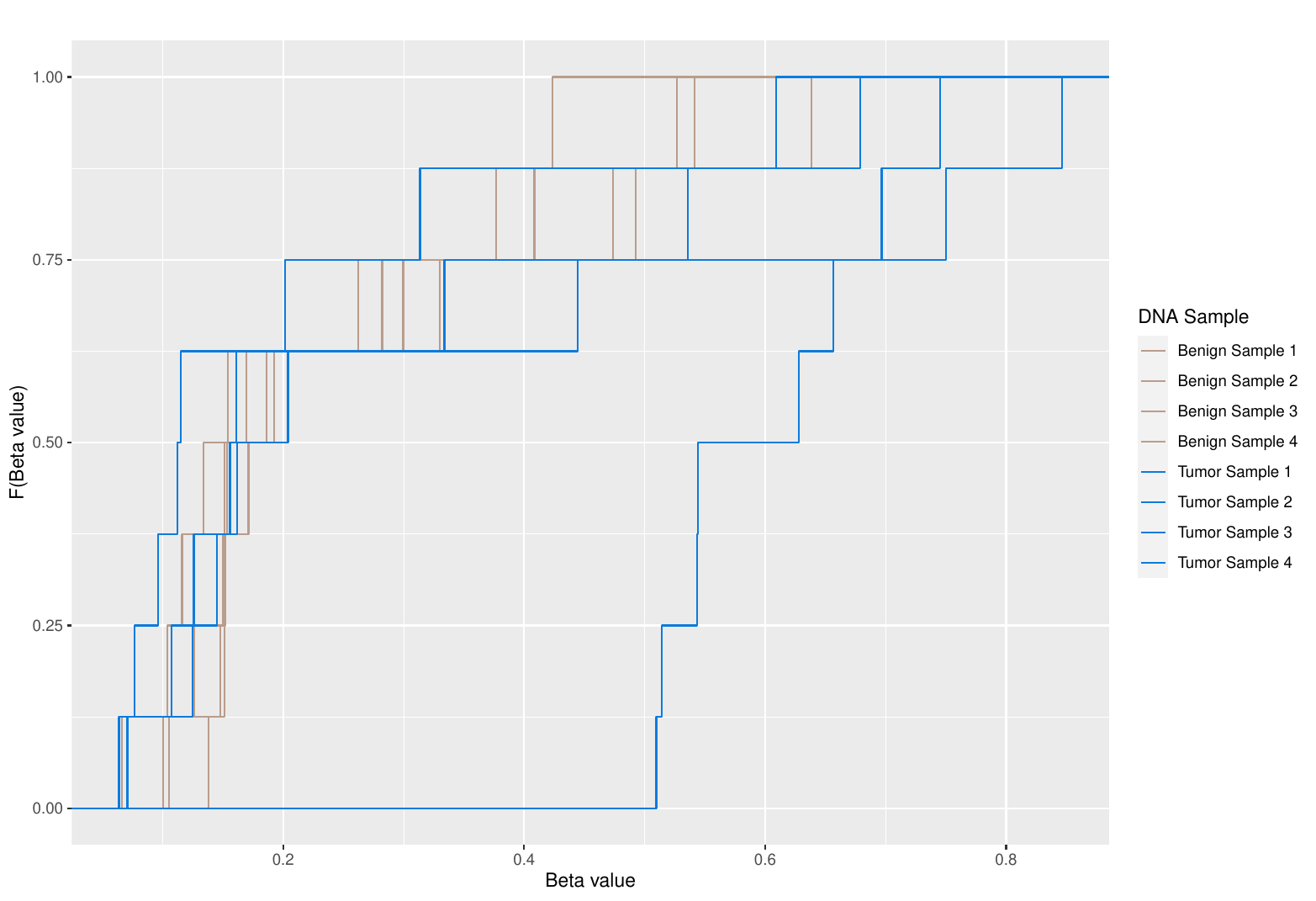}
\caption{ECDFs for all the CpG sites identified as  mostly differentially methylated and related to the GPX3 genes for all patient samples.}
\label{ESCC:Figure6}
\end{center}
\end{figure*}

We conducted gene ontology analysis \citep{gometh} on the differentially methylated clusters, namely cluster 1 and cluster 2. In cluster 1, a total of 439 significant biological processes were identified (FDR adjusted p-value $< 0.05$), while cluster 2 revealed 385 significant biological processes. Additionally, the DMCs within cluster 1 exhibited relevance to 37 significant KEGG pathways, while those within cluster 2 were associated with 26 significant KEGG pathways.
The K$\cdot$R model identified DMCs related to genes implicated in esophageal squamous cell carcinogenesis. For example, the expression of the GPX3 gene has been shown to be downregulated in ESCC when compared with normal esophageal mucosa \citep{gpx3}. The promoter methylation results in the silencing of the GPX3 genes in ESCC. The ECDF plot in Figure \ref{ESCC:Figure6} illustrates hypermethylation of the identified DMCs related to the GPX3 gene in the tumour samples.

\null \pagebreak

\end{document}